\documentclass[pre]{revtex4}
\usepackage{amsmath,amssymb}
\usepackage{epsfig}

\begin{document}

\newlength{\mylen}
\setlength{\mylen}{\textwidth}
\addtolength{\mylen}{-1cm}
\newcommand{\bea}{\begin{eqnarray}}
\newcommand{\eea}{\end{eqnarray}}
\newcommand{\spaceint}[2]{\int_{#1} d^3 #2 \;}
\newcommand{\vect}[1]{\mathbf{#1}}
\newcommand{\vat}{V^{\rm att}}


\title{Effective capillary interaction of spherical particles at  fluid interfaces}

\author{M. Oettel}
\author{A. Dom\'\i nguez}
\author{S. Dietrich}
\affiliation{Max--Planck--Institut f\"ur Metallforschung, Heisenbergstr.~3, D--70569 Stuttgart, Germany,}
\affiliation{Institut f\"ur Theoretische und Angewandte
  Physik, Universit\"at Stuttgart, Pfaffenwaldring 57, D--70569 Stuttgart, Germany}

\date{\today}

\begin{abstract}
  
  We present a detailed analysis of the effective force between two
  smooth spherical colloids floating at a fluid interface due to
  deformations of the interface. The results hold in  general and are
  applicable independently of the source of the deformation provided
  the capillary deformations are small so that a
  superposition approximation for the deformations is valid. We
  conclude that an effective long--ranged attraction is possible if the
  net force on the system does not vanish. Otherwise, the interaction
  is short--ranged and cannot be computed reliably based on the
  superposition approximation.
  As an application, we consider the case of like--charged, smooth
  nanoparticles and electrostatically induced capillary
  deformation. The resulting long--ranged capillary attraction can be easily
  tuned by a relatively small external electrostatic field, but it
  cannot explain recent experimental observations of attraction if these 
  experimental systems were indeed isolated. 

\end{abstract}

\maketitle 

\section{Introduction}

In view of various applications such as the study of two--dimensional melting
\cite{Pie80}, investigations of mesoscale structure formation \cite{Joa01}
or engineering of colloidal crystals on spherical surfaces \cite{Din02}, 
the self--assembly of 
sub-$\mu$m colloidal particles at water--air or water--oil interfaces
has gained much interest in recent years. These particles are trapped at the interface
 if the colloid is only partially wetted by both the water and the oil. 
This configuration is stable against thermal fluctuations and appears to be the global equilibrium state,
as
it is observed experimentally that the colloids immersed in the bulk phases
are attracted to the interface \cite{Pie80} (see also Sec.~\ref{sec:1coll}).
The mutual interaction between the trapped colloids at distances close to contact,
i.e., within the range of molecular forces,  
is dominated by strong van--der--Waals attraction. In order to avoid coagulation
due to 
this attraction, the colloids can be stabilized sterically with polymers  
or with charges such that the colloids repel each other. Variants of charge 
stabilization may include the coverage with ionizable molecules which dissociate 
in water, or the labeling of colloids with charged fluorescent markers.
For charge--stabilized colloids at large distances, the resulting repulsive force 
at water interfaces
stems from a dipole--dipole interaction as shown theoretically
for point charges on surfaces \cite{Hur85} and verified experimentally for 
polystyrene (PS) spheres on water--oil interfaces \cite{Ave02}.   

Nonetheless, charged colloids at interfaces also show attractions far beyond the
range of van--der--Waals forces. The corresponding experimental evidence can be roughly 
classified as follows. (i) According to 
Refs.~\cite{Ghe97,Ghe01,Gar98a,Gar98b,Sta00}, PS spheres 
(radii $R=0.25\dots 2.5$ $\mu$m)
 on flat water--air interfaces  using highly deionized water exhibit  
spontaneous formation
of complicated metastable mesostructures. They are consistent with the presence
of an attractive, secondary  
minimum in the intercolloidal potential at distances $d/R\approx 3\dots 10$ with
a depth of a few $k_B T$. The use of water slightly contaminated by ions seems to 
move the minimum further out and to reduce its depth \cite{Sta00,Que01}. 
(ii) In Ref.~\cite{Nik02}, PMMA spheres with radius $R=0.75$ $\mu$m were trapped
at the interface of water droplets immersed in oil. 
%
Here, the secondary minimum has been measured at a 
distance $d/R=7.6$ and is reported to be  surprisingly steep. 
The tentative explanation of these
findings given in Ref.~\cite{Nik02} invokes an analogue of long--ranged
flotation or capillary forces which decay $\propto 1/d$. 
This interpretation  was criticized in 
Ref.~\cite{Meg03} (with which the authors of Ref.~\cite{Nik02} agreed \cite{Nik03})
and in Ref.~\cite{For04}
which both concluded that possible capillary forces in
this system are much shorter ranged, i.e.,   $\propto d^{-7}$, 
but the authors of these references disagree with respect to
the sign of this shorter--ranged force. In yet another twist of the story, 
after completion of our work we encountered the very recent Ref.~\cite{Kra04}
in which the authors claim that long--ranged
capillary forces $\propto 1/d$ caused by the colloidal charges persist
for sub-$\mu$m particles.
This conclusion is based on measurements of the meniscus shape around single glass
spheres with radii 200 \dots 300 $\mu$m floating at water--oil and water--air interfaces.

Motivated by the experimental data summarized above  and the still incomplete 
theoretical understanding, here we
undertake a quite general analysis of capillary interactions
between two spherical colloids trapped at fluid interfaces. We characterize the system 
by a general stress field which acts on the interface, e.g.,  due
to  a discontinuous electrostatic field at the interface, 
and by forces on the colloid of  
gravitational and/or electrostatic origin. Special attention will be given to the role
of a restoring force fixing the interface, e.g., due to gravity or interface pinning.
In Sec.~\ref{sec:1coll} we present a free energy model for a single
colloid trapped at an interface in the limit of small stresses and forces. The
general solution for the interface deformation will be used in 
Sec.~\ref{sec:2coll} in order  to determine the effective potential 
between two colloids within the 
superposition approximation. In view of the differing theoretical results in the literature,
the derivation of the interface deformation and the resulting effective potential 
is presented in detail in order to reveal
properly the subtleties involved. 
In Sec.~\ref{sec:discussion} we apply the results 
to the case of charged polymeric spheres on water interfaces. 
It turns out that the constraint of approximate mechanical isolation
of the experimental systems renders the capillary interaction basically short--ranged. 
Long--ranged attractive forces $\propto 1/d$  only arise through a restoring force 
acting on the interface
(which is, however, weak) or in the presence of an external electric field. 
We shall discuss the relation to previous theoretical results, especially
in view of the experimental results reported in Refs.~\cite{Nik02,Kra04}.
Directions for further research will be pointed out. In Sec.~\ref{sec:summary} we summarize our
results.

\section{Equilibrium state of a single colloid}
\label{sec:1coll}

\begin{figure}
  \begin{center}
    \epsfig{file=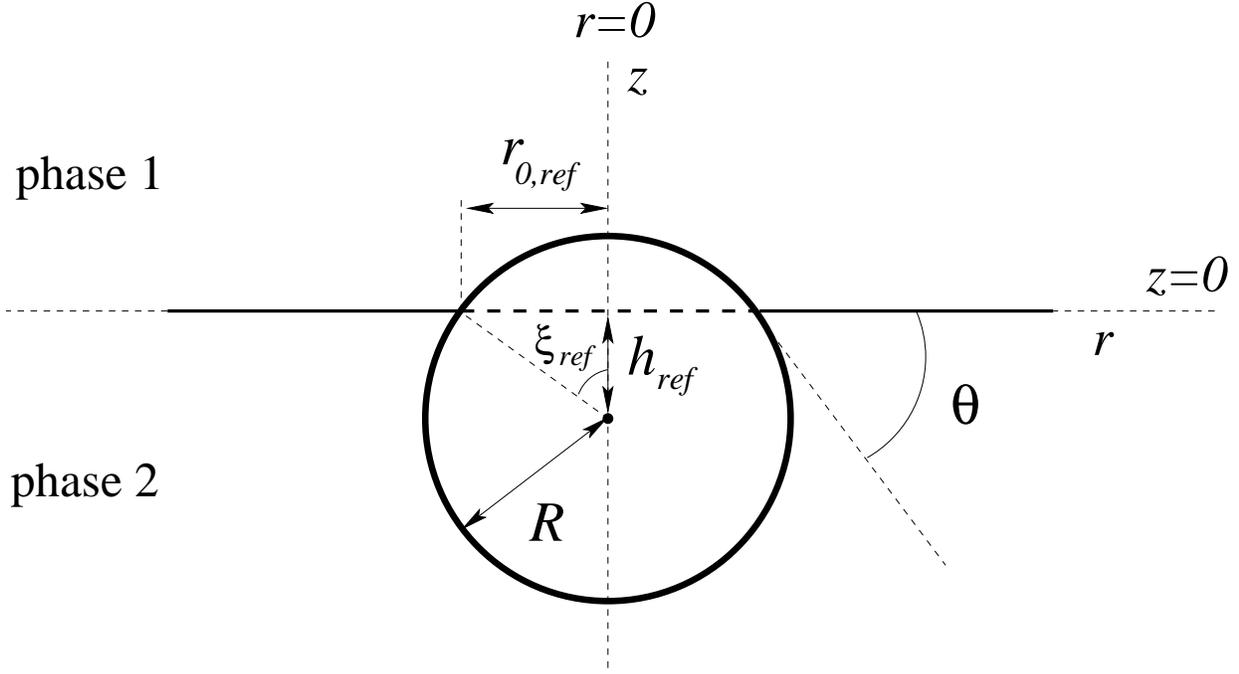,width=.95\textwidth}
    \caption{
      Geometry of the reference state. The equilibrium contact angle
      $\theta$ fixes the height $h_{\rm ref} = - R \cos \theta$ of
      the colloid center, the contact radius $r_{0, \rm ref} = R
      \sin \theta$, and the auxiliary angular variable $\xi_{\rm ref}=\theta$.}
    \label{fig:ref}
  \end{center}
\end{figure}

\begin{figure}
  \begin{center}
    \epsfig{file=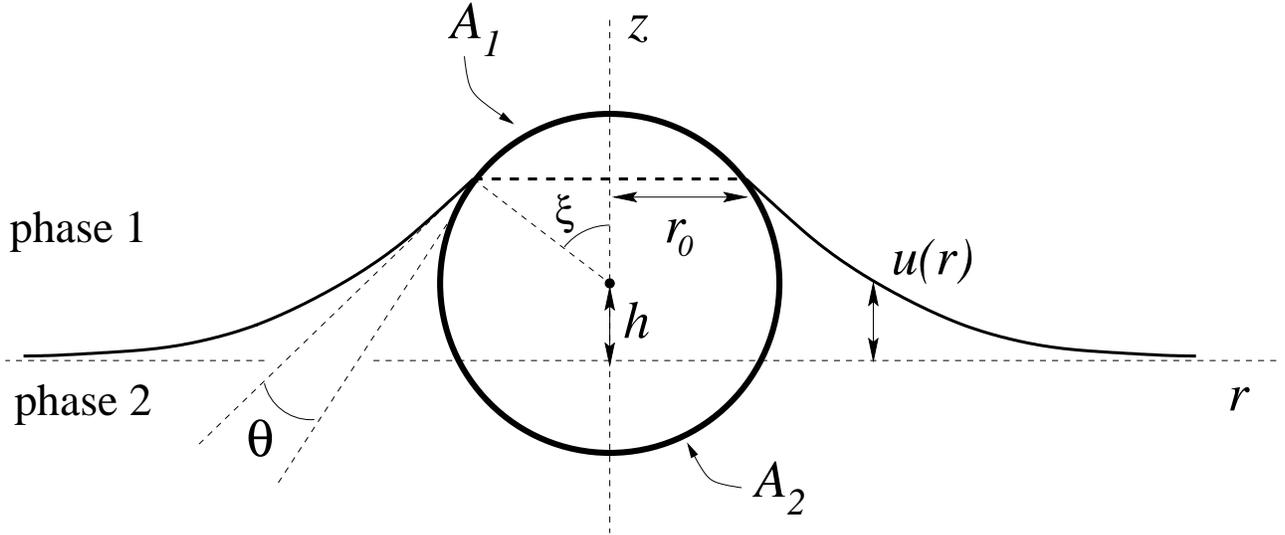,width=.95\textwidth}
    \caption{
      Description of deviations from the reference state. $u(r)$ is
      the meniscus profile and $h$ is the height of the colloid
      center. The angle $\xi$ and the radius $r_0 = R \sin
      \xi$ of the three--phase contact line 
      are auxiliary variables, which depend on $u(r)$ and $h$
      through the geometrical relationship $h = u(r_0) - R \cos
      \xi$. In this mesoscopic description $r_0$ is defined as the
      position where the meniscus profile intersects the surface of the
      sphere. $A_i$ is the surface area of the colloid exposed to
      phase $i$.}
    \label{fig:pert}
  \end{center}
\end{figure}

In this section we consider as a first step the equilibrium state of a
single colloid of radius $R$ at the interface between two
fluid phases denoted as 1 and 2. 
The contact angle formed by the interface and the colloid surface is
given by Young's law,
\begin{equation}
  \label{eq:young}
  \cos \theta = \frac{\gamma_1 - \gamma_2}{\gamma} ,
\end{equation}
where $\gamma$ is the surface tension between phases 1 and 2, and
$\gamma_i$ is the surface tension between the colloid and phase $i$.
As a reference state, with respect to which changes in free energy will
be measured, we take a planar meniscus configuration with the colloid
at such a height $h_{\rm ref}$ that Young's law is satisfied
(Fig.~\ref{fig:ref}).  This corresponds to the equilibrium
configuration of an uncharged colloid at the interface if its weight
can be neglected --- which for generic cases is a safe approximation
for $R \lesssim 1$ $\mu$m \cite{Kra00} (see also Sec.~\ref{sec:discussion}).
We model the colloid as a smooth sphere so that the system is
invariant under rotations around the colloid axis perpendicular to the
reference planar meniscus.
The presence of charges induces a shift of the system (colloid and interface) with
respect to the reference state. Here we neglect corresponding
  changes in the surface tensions $\gamma$ and $\gamma_i$; this
  approximation is expected to be valid provided the concentration of
  charges is sufficiently small. This shift is characterized by the
meniscus profile $u(r)$ relative to the planar configuration and by
the height $h$ of the colloid center. In the reference configuration, 
the charge distribution is assumed to be already in equilibrium.
In the following we do not consider the degree of freedom ``charge density field'' 
explicitly but take it to be fixed to that of the reference configuration.
This amounts to neglecting the feedback of the interface displacement on
the charge distribution.
It turns out to be useful to
introduce the radius $r_0$ of the three--phase contact line and the angle $\xi$ as auxiliary
variables (see Fig.~\ref{fig:pert}).

\subsection{The free energy}

In this subsection we formulate a free energy functional for the
degrees of freedom $u(r)$ and $h$. As shown later, for the cases of
interest here the deviations from the reference configuration are
small enough to justify a perturbative treatment, and the free energy
can be expanded up
to quadratic order in $\Delta h:=h-h_{\rm ref}$ 
and $u(r)$.
We denote with $\Pi(r)$ the vertical force per unit area acting on the
meniscus surface in the reference configuration. In the case of a
charged colloid, $\Pi(r)$ is given by the $zz$--component of the
difference of the Maxwell stress tensor right above and below the
meniscus, plus the pressure difference acting across the meniscus (incluiding an
imbalance in osmotic pressure due to the different concentration of
ions just above and below the meniscus) \cite{For04,SMN02}. We introduce the
following dimensionless parameter to measure the relative strength of
this force:
\begin{equation}
  \label{eq:epsPi}
  \varepsilon_\Pi := \frac{1}{2 \pi \gamma r_{0,\rm ref}} \int_{S_{\rm men,ref}} \!\!\!\!\!\! dA\; \Pi =
  \frac{1}{\gamma r_{0,\rm ref}} \int_{r_{0,\rm ref}}^\infty dr\; r\,\Pi(r) ,
\end{equation}
where the integral extends over the flat meniscus $S_{\rm men,ref}$
 (the plane $z=0$
with a circular hole of radius $r_{0, \rm ref}$). We assume
$\Pi(r \to \infty) \sim r^{-n}$ with $n>2$, so that the integral
converges.  
In the same spirit we introduce
the total vertical force ${\bf F}= F{\bf e}_z$ acting on the charged colloid in the flat
reference configuration, which includes gravity, the electrostatic
force, and the total (i.e., hydrostatic and osmotic) pressure exerted
by the fluids. This leads to the definition
\begin{equation}
  \label{eq:epsF}
  \varepsilon_F := - \frac{F}{2 \pi \gamma r_{0,\rm ref}} .
\end{equation}
These two dimensionless parameters will appear naturally in the course
of the calculations. If $\varepsilon_\Pi$ and $\varepsilon_F$ vanish,
the reference configuration {\em is} the equilibrium state; 
it is the global minimum of the free energy functional given in
Eqs.~(\ref{eq:Fdef}) and (\ref{eq:F}) below if $\Pi \equiv 0$, $F \equiv 0$.
The
aforementioned perturbative expansion of the free energy can be
rephrased as an expansion in terms of the small parameters $\varepsilon_\Pi$ and 
$\varepsilon_F$.

The free energy ${\cal F}$ of the colloid expressed relative to the
aforementioned reference configuration consists of five terms: 
\begin{equation}
  \label{eq:Fdef}
  {\cal F} = {\cal F}_{\rm cont} + {\cal F}_{\rm men} + 
  {\cal F}_{\rm vol} + {\cal F}_{\rm inter} + {\cal F}_{\rm coll} .
\end{equation}
In the following we discuss each contribution:
%
\begin{itemize}
\item {\it Fluid contact of the colloid}. With $A_i$ denoting the surface
  area of the colloid which is in contact with phase $i$, the surface
  free energy of the colloid due to its exposure to the phases 1 and 2
  is
  \begin{equation}
    \label{eq:Fcont}
    {\cal F}_{\rm cont} = \gamma_1 A_1 + \gamma_2 A_2 
    - (\gamma_1 A_{1, \rm ref} + \gamma_2 A_{2, \rm ref} ) .
  \end{equation}
In Appendix~\ref{sec:Fcont} we express this contribution as a function
  of $\Delta h$ and $u(r)$. The final result, valid up to corrections
  of at least third order in $\varepsilon_\Pi$ or $\varepsilon_F$,
  reads
  \begin{equation}
    \label{eq:Fcont_iso}
    {\cal F}_{\rm cont} \simeq \pi \gamma [ u(r_{0, \rm ref}) - \Delta
    h ]^2 + \pi \gamma (r_0^2 - r_{0,\rm ref}^2) .
  \end{equation}
  
\item {\it Change of the meniscus area}. The free energy contribution
  due to variations in the meniscus area relative to the reference
  state reads
  \begin{equation}
    {\cal F}_{\rm men} = \gamma \int_{S_{\rm men}} \!\!\!\!\!\! dA \; \sqrt{1+|\nabla u|^2} - 
    \gamma \int_{S_{\rm men,ref}} \!\!\!\!\!\! dA ,
  \end{equation}
  where $S_{\rm men}$ is the surface of the fluid interface projected
  onto the plane $z=0$ (in which the reference interface is located), 
  and $\nabla$ is the gradient operator on the flat reference interface.
  %
  For small slopes ($|\nabla u| \ll 1$) one  obtains:
  \begin{eqnarray}
    \label{eq:Fmen}
    {\cal F}_{\rm men} & \simeq & \gamma \int_{S_{\rm men}\backslash S_{\rm men,ref}} \!\!\!\!\!\! dA +
    \frac{1}{2} \gamma \int_{S_{\rm men}} \!\!\!\!\!\! dA \; |\nabla
  u|^2 \nonumber \\
        & \simeq & \pi \gamma (r_{0,\rm ref}^2 - r_0^2) +
    \frac{1}{2} \gamma \int_{S_{\rm men,ref}} \!\!\!\!\!\! dA \; |\nabla u|^2 \;.
  \end{eqnarray}
 Since the $u$--dependent term is of second order in $u$,
  we have approximated the integration domain $S_{\rm men}$ by $S_{\rm men, ref}$; 
  the corrections are at least of third
  order in the small parameters $\varepsilon_\Pi$ or $\varepsilon_F$.
The first term in this expression represents the change in the area of
  the meniscus which is cut out by the colloid, and in Eq.~(\ref{eq:Fdef}) 
 it cancels 
  the second term of Eq.~(\ref{eq:Fcont_iso}).

\item {\it Volume forces on the fluids}. We consider the case that the
only volume force acting on the fluid phases is gravity. The
electrostatic forces are active only at the surfaces, where a net
charge can accumulate. 
%
%
If the spherical colloid is replaced by a long
cylinder of radius $r_{0, \rm ref}$ (Fig.~\ref{fig:cylinder}) the
change in gravitational potential energy relative to the reference
state due to the displacements of volumes of the fluid phases can be
calculated easily: 
\begin{equation} 
\label{eq:Fvol}
{\cal F}_{\rm vol} = \frac{1}{2} \gamma \int_{S_{\rm men,ref}}
\!\!\!\!\!\! dA \; \frac{u^2(r)}{\lambda^2} = \pi \gamma \int_{r_{0,
\rm ref}}^\infty dr \; r \, \frac{u^2(r)}{\lambda^2} \, , 
\end{equation}
assuming that the mass density of phase 2 is larger than the one of
phase 1, i.e., $\varrho_2 > \varrho_1$; $\lambda$ denotes the
capillary length
\begin{equation}
\lambda=\sqrt{\frac{\gamma}{|\rho_2-\rho_1|g}} 
\end{equation} 
in terms of the acceleration $g$ of gravity. For a spherical colloid ${\cal
F}_{\rm vol}$ in Eq.~(\ref{eq:Fvol}) has to be corrected to account
for the specific dependence of the displaced fluid volume on the colloid
height $h$,
yielding a slightly cumbersome expression \cite{PKDN93}.  This correction is
usually numerically small and we neglect it for reasons of
simplicity. As it will be shown in Subsec.~\ref{sec:lambda} this is
justified because the results of interest here are insensitive to the
precise form of ${\cal F}_{\rm vol}$ in the limit $\lambda \rightarrow
\infty$ 
($\lambda \approx 1$ mm, which is much larger than any other relevant
length scale of the system).  
\begin{figure} 
\begin{center}
\epsfig{file=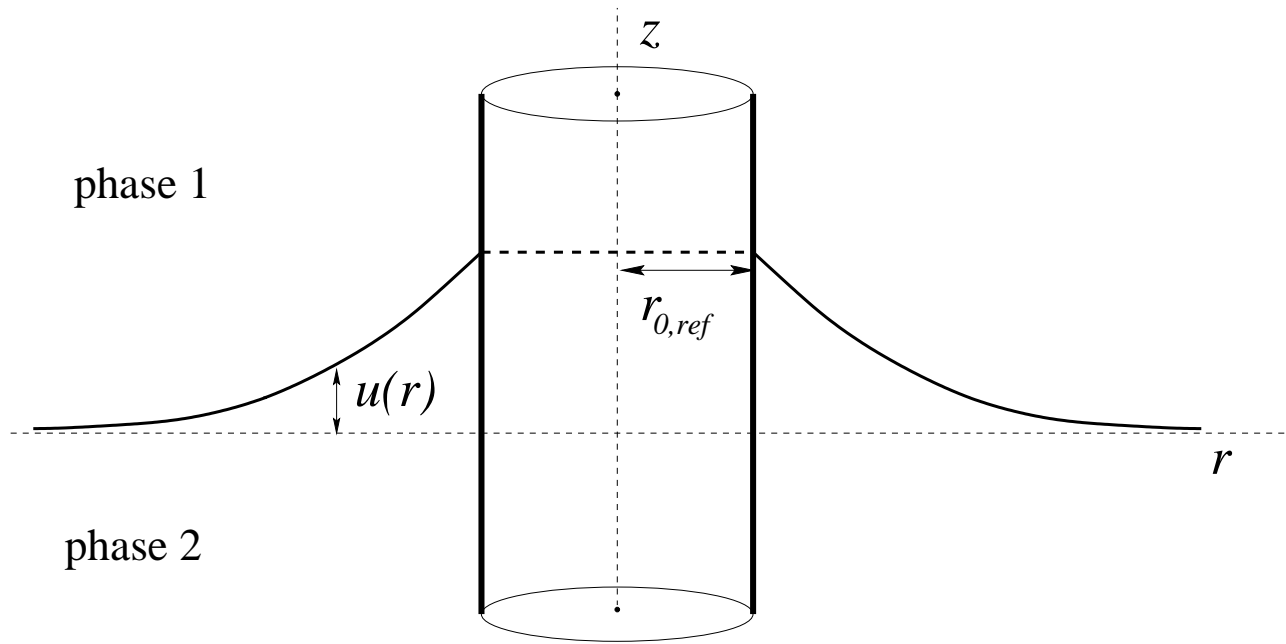,width=.95\textwidth} 
\caption{
Long cylinder of radius $r_{0, \rm ref}$ immersed
vertically into the fluid interface between the phases 1 and 2.}
\label{fig:cylinder} 
\end{center} 
\end{figure}
  
\item {\it Force on the fluid interface}. The aforementioned surface
force density $\Pi(r)$ acts on the fluid interface between phase 1 and
phase 2.  The free energy change due to the ensuing displacements of
the meniscus is 
\begin{equation} \label{eq:Finter} {\cal F}_{\rm
inter} \simeq - \int_{S_{\rm men}} \!\!\!\!\!\! dA \; \Pi u \simeq -
\int_{S_{\rm men, ref}} \!\!\!\!\!\! dA \; \Pi u = - 2 \pi \int_{r_{0,
\rm ref}}^\infty dr \; r \, \Pi(r) u(r) .  \end{equation} 
Here
$\Pi(r)$ is the surface force in the {\em reference} configuration
($\Pi>0$ corresponds to a force pointing upward).  Changes in the
force induced by meniscus deformations and colloidal displacements
contribute terms of higher orders in $\varepsilon_\Pi$ or
$\varepsilon_F$ in Eq.~(\ref{eq:Finter}). The replacement of $S_{\rm
men}$ by $S_{\rm men,ref}$ in Eq.~(\ref{eq:Finter}) introduces terms
of higher order, too.
 
\item {\it Contribution from the colloid}. The free energy change due to
  a vertical displacement of the colloid is
  \begin{equation}
    \label{eq:Fcoll}
    {\cal F}_{\rm coll} \simeq - F \Delta h, 
  \end{equation}
  where $F$ is the vertical force on the colloid in the {\em reference}
    configuration. 
Like for ${\cal F}_{\rm inter}$, 
  changes of $F$ due to deviations from the reference configuration contribute to higher order terms.
\end{itemize}

In conclusion, by adding Eqs.~(\ref{eq:Fcont_iso}), (\ref{eq:Fmen}),
(\ref{eq:Fvol}), (\ref{eq:Finter}), and (\ref{eq:Fcoll}) we obtain the
following approximate expression for the total free energy, which is correct up
to second order in $\varepsilon_\Pi$ or $\varepsilon_F$, and which is a
function of $\Delta h$ and a functional of $u(r)$:
\begin{equation}
  \label{eq:F}
  {\cal F} \simeq 
  2 \pi \gamma \int_{r_{\rm 0, ref}}^{\infty} dr \; r \left[ \frac{1}{2}\left(\frac{d u}{d r} \right)^2 +
    \frac{u^2}{2 \lambda^2} - \frac{1}{\gamma} \Pi \, u \right] +
  \pi \gamma [ u_0 - \Delta h ]^2 -
  F \Delta h ,
\end{equation}
where $u_0 \equiv u(r_{0, \rm ref})$.

\subsection{Minimization of the free energy}
\label{sec:minF}

The equilibrium configuration of the system minimizes the free
energy expression~(\ref{eq:F}). The minimization proceeds in two stages.  First
we seek the minimum with respect to $\Delta h$ at fixed $u(r)$:
\begin{equation}
  \label{eq:h_eq}
  \frac{\partial {\cal F}}{\partial (\Delta h)} = 0 \qquad \Rightarrow \qquad 
  \Delta h = u_0 - \varepsilon_F r_{0, \rm ref} ,
\end{equation}
where we have used the definition~(\ref{eq:epsF}). Using repeatedly
the definitions of $h$ and $r_0$ in terms of the auxiliary angle
$\xi$ (Fig.~\ref{fig:pert}) we can compute the change of the contact
radius:
\begin{equation}
  \label{eq:r_0,eq}
  \Delta r_0 \equiv r_0 - r_{0, \rm ref} = R(\sin \xi - \sin \theta) 
  \simeq (\theta - \xi)  h_{\rm ref} 
  \simeq \frac{\Delta h - u_0}{r_{0, \rm ref}}  h_{\rm ref} 
  = - \varepsilon_F h_{\rm ref} ,
\end{equation}
plus corrections of second order in $\varepsilon_\Pi$,
$\varepsilon_F$.
In the second step, we minimize Eq.~(\ref{eq:F}) with respect to
$u(r)$ at fixed $\Delta h$. This is a problem of variations with a
free boundary condition at $r=r_{0, \rm ref}$ \cite{CoHi}. Variation
with respect to $u(r \neq r_{0, \rm ref})$ yields a second--order
ordinary differential equation,
\begin{equation}
  \label{eq:younglaplace}
  \frac{d^2 u}{d r^2} + \frac{1}{r} \frac{d u}{d r} - 
  \frac{1}{\lambda^2} u = - \frac{1}{\gamma} \Pi (r) ,
\end{equation}
while variation with respect to $u_0$ provides a boundary condition:
\begin{equation}
  \label{eq:u'0}
  u'_0 := \left. \frac{d u}{d r}\right|_{r=r_{0, \rm ref}} =  
  \frac{u_0 - \Delta h}{r_{0, \rm ref}} = \varepsilon_F ,
\end{equation}
where the last equality follows from Eq.~(\ref{eq:h_eq}).
The second boundary condition one has to impose on
Eq.~(\ref{eq:younglaplace}) follows from the requirement that the
meniscus is asymptotically flat far from the colloid (assuming that
$\Pi(r \to \infty)=0$):
\begin{equation}
  \label{eq:u_infty}
  \lim_{r \to \infty} u(r) = 0 . 
\end{equation}
Equation~(\ref{eq:younglaplace}) describes mechanical equilibrium of the
interface such that the Laplace pressure balances the forces acting on
the interface. The boundary condition~(\ref{eq:u'0}) expresses
mechanical equilibrium of the colloidal particle: 
At the contact line (i.e., the circle with radius $r_0$ at $z=u_0$) the interface
exerts a force onto the colloid which has a non--vanishing contribution only
in $z$--direction with the magnitude $2 \pi \gamma r_{0} \sin[\arctan(u_0')]
\approx 2 \pi \gamma r_{0,\rm ref} u'_0$. This contact line force is balanced by the
total force $F$.

The solution of Eqs.~(\ref{eq:younglaplace}--\ref{eq:u_infty}) can be
written in terms of the modified Bessel functions of zeroth order: 
\begin{equation}
  \label{eq:solBessel}
  u(r) = \frac{1}{\gamma} I_0 \left(\frac{r}{\lambda}\right) \int_r^{\infty} \!\!\! ds \; s\, \Pi(s) 
  K_0 \left(\frac{s}{\lambda}\right) +
  \frac{1}{\gamma} K_0 \left(\frac{r}{\lambda}\right) \left[ A + 
    \int_{r_{0, \rm ref}}^r \!\!\! ds \; s\, \Pi(s) I_0\left(\frac{s}{\lambda}\right) \right] ,
\end{equation}
where the integration constant $A$ is determined by the boundary
condition~(\ref{eq:u'0}).

\subsection{Asymptotic behavior in the limit $\lambda \to \infty$}
\label{sec:lambda}

For typical values of the parameters, $\lambda$ is of the order of
millimeters and therefore much larger than any other length scale
occurring for experiments with 
sub-micrometer colloids. In order to study the intermediate
asymptotics ($r_{0, \rm ref}, r \ll \lambda$) of $u(r)$ as given by
Eq.~(\ref{eq:solBessel}), we insert the asymptotic expansions of the
Bessel functions \cite{AbSt} as $\lambda \to \infty$ and retain those
terms which do not vanish in this limit.  Assuming that $\Pi(r \to
\infty)$ decays sufficiently fast, Eq.~(\ref{eq:solBessel}) reduces to
\begin{equation}
  \label{eq:sol1}
  u (r) \simeq r_{\rm 0, ref} (\varepsilon_\Pi - \varepsilon_F) \ln\frac{C \lambda}{r}
  - \frac{1}{\gamma} \int_{r}^{+\infty} \!\!\!\! ds \; s \, \Pi(s) \ln \frac{s}{r} ,
\end{equation}
where $C = 2 e^{-\gamma_{\rm E}} \simeq 1.12$ and $\gamma_{\rm E}$ is
Euler's constant. In Eq.~(\ref{eq:sol1}), we have expressed the
integration constant $A$ appearing in Eq.~(\ref{eq:solBessel}) in
terms of $\varepsilon_F$ by using the boundary condition~(\ref{eq:u'0}). The
first term in Eq.~(\ref{eq:sol1}) is a solution of the homogeneous part
of Eq.~(\ref{eq:younglaplace}) (with $\lambda^{-1}=0$ in the
equation), demonstrating that the limit $\lambda \to \infty$ is
singular as long as $\varepsilon_\Pi \neq \varepsilon_F$.
The second term corresponds to a particular solution of the
inhomogeneous differential equation. If the surface force $\Pi(r)$
decays algebraically, $\Pi(r \to \infty) \propto r^{-n}$, this term
decays like $r^{2-n}$ (since we have assumed $n>2$), so that the logarithmic
contribution is dominant. If $n \leq 2$, the asymptotic behavior is no
longer given by Eq.~(\ref{eq:sol1}) but a different dependence on
$\lambda$ arises.

\begin{figure}
    \begin{center}
      \epsfig{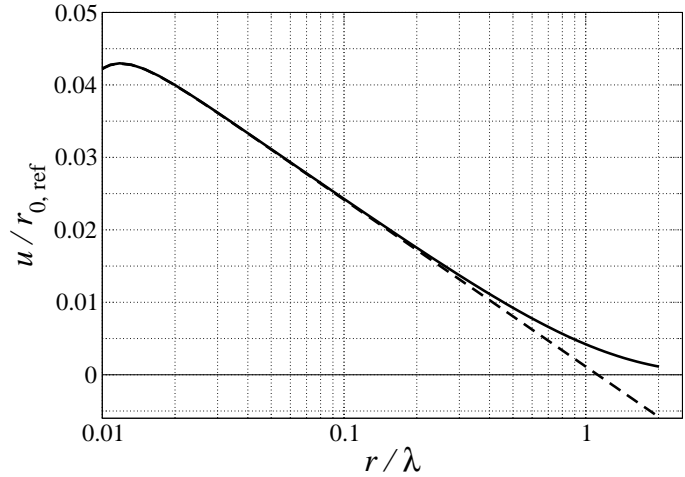}
      \caption{
        The meniscus solution for the parameter choice $\lambda/r_{0,\rm ref}= 10^2$,
        $\varepsilon_F=10^{-2}$, and a dipole--like stress field
        $r_{0,\rm ref}\:\Pi(r) / \gamma = 0.08\:(r_{0,\rm ref}/r)^6$ 
        ($\varepsilon_\Pi=2\cdot 10^{-2}$).
        The solid line represents the solution $u(r)$ given by
        Eq.~(\ref{eq:solBessel}). The dashed line is the intermediate asymptotic
        solution given by Eq.~(\ref{eq:sol1}). 
        Note that the capillary length $\lambda$ is typically of $O$(1 mm). In
        the present context we focus on the length scale $r_{0,\rm ref},\:
        r \ll \lambda$, for which Eq.~(\ref{eq:sol1}) holds.  }
      \label{fig:u(r)}
    \end{center}
  \end{figure}

At distances $r$ of the order of $\lambda$, the
expression~(\ref{eq:sol1}) is not valid and a crossover to the exact
solution~(\ref{eq:solBessel}) takes place in order to satisfy the
boundary condition at infinity
(Eq.~(\ref{eq:u_infty})). Figure \ref{fig:u(r)} sketches the behavior of
the meniscus profile $u(r)$. As can
be checked directly in the differential
equation~(\ref{eq:younglaplace}), 
one has $u(r) \sim (\lambda^2/\gamma) \Pi(r) \propto r^{-n}$ asymptotically as $r \to +\infty$, 
which expresses the balance between
the surface force $\Pi(r)$ on the meniscus and the gravitational
force. There is, however, another intermediate regime $r \gg \lambda$
but not too large, in which $u(r)$ decays $\propto \exp(-r/\lambda)$ and which
corresponds to the solution of Eq.~(\ref{eq:younglaplace}) with
$\Pi$ set to zero, i.e., when gravity is balanced by the Laplace
pressure induced by the meniscus curvature. 

These conclusions, as well as the functional dependence of $u(r)$ on
$\Pi(r)$, are robust and independent of the details of the boundary
condition at $r \to \infty$. In order to support this statement, we
consider two cases which implement the distant boundary condition differently:
\begin{itemize}
\item {\it Pinned interface}. In the absence of gravity the interface
  is assumed to be pinned at a finite distance $L$ from the colloid.
  This corresponds to setting $\lambda^{-1}=0$ in
  Eq.~(\ref{eq:younglaplace}) and replacing Eq.~(\ref{eq:u_infty}) by
  the new boundary condition $u(L)=0$. In this case the solution of
  Eq.~(\ref{eq:younglaplace}) is given by
  \begin{equation}
    \label{eq:sol_pin}
    u (r) = r_{\rm 0, ref} (\tilde{\varepsilon}_\Pi - \varepsilon_F) \ln\frac{L}{r}
    - \frac{1}{\gamma} \int_{r}^{L} \!\!\!\! ds \; s \, \Pi(s) \ln \frac{s}{r} ,
  \end{equation}
  where
  \begin{equation}
    \label{eq:epsPi_pin}
    \tilde{\varepsilon}_\Pi := \frac{1}{\gamma r_{0,\rm ref}} \int_{r_{0,\rm ref}}^L dr\; r\,\Pi(r) ,
  \end{equation}
  in analogy to Eq.~(\ref{eq:epsPi}). In the intermediate asymptotic
  regime ($r_{0, \rm ref},\: r \ll L$), the meniscus profile is then given by
  \begin{equation}
    \label{eq:sol2}
    u (r) \simeq r_{\rm 0, ref} (\varepsilon_\Pi - \varepsilon_F) \ln\frac{L}{r}
    - \frac{1}{\gamma} \int_{r}^{+\infty} \!\!\!\! ds \; s \, \Pi(s) \ln \frac{s}{r} ,
  \end{equation}
  assuming that $\Pi(r)$ decays sufficiently fast so that the
  integrals in Eqs.~(\ref{eq:sol_pin}) and (\ref{eq:epsPi_pin}) converge as
  $L \to \infty$.  Equation (\ref{eq:sol2}) resembles Eq.~(\ref{eq:sol1})
  with $C \lambda$ replaced by $L$.
  
\item {\it Pinned curved reference interface}. In some experiments the
  interface between phases 1 and 2 is in fact closed, so that the
  colloidal particle lies at the surface of a large nonvolatile spherical droplet of
  phase 2 which is immersed in phase 1 \cite{Nik02} (Fig.~\ref{fig:droplet}) 
%
  and  fixed by certain means (e.g.~by a glass plate).
  \begin{figure}
    \begin{center}
      \epsfig{file=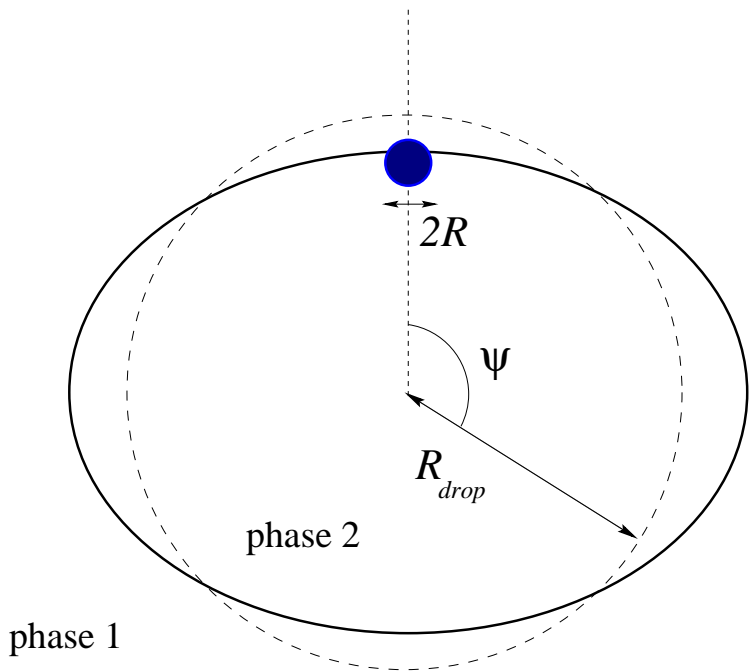,width=.55\textwidth}
      \caption{
        Colloid at the surface of a droplet of phase 2 immersed into
        phase 1. The radius $R_{\rm drop}$ of the droplet, which is
        spherical without the colloid, is usually much
        larger than the colloid radius $R$.
        (The deformation of the droplet has been exaggerated.)}
      \label{fig:droplet}
    \end{center}
  \end{figure}
  The free energy functional~(\ref{eq:F}) has to be modified to
  account for the curvature of the reference interface as well as of
  the constraint that the droplet volume remains unchanged under
  deformation.
  To determine the interface deformation, we minimize the functional
  and employ the boundary condition that the droplet is fixed at
  some point far from the colloid. The mathematical details and the
  corresponding solution for $u(r)$ are presented in
  Appendix~\ref{sec:droplet}.  Here we quote only the intermediate
  asymptotic behavior ($r_{0, \rm ref},\; r \ll R_{\rm drop}$, where
  $R_{\rm drop}$ is the radius of the undeformed droplet):
  \begin{equation}
    \label{eq:sol3}
    u (r) \simeq r_{\rm 0, ref} (\varepsilon_\Pi - \varepsilon_F) \ln\frac{\tilde{C} R_{\rm drop}}{r}
    - \frac{1}{\gamma} \int_{r}^{+\infty} \!\!\!\! ds \; s \, \Pi(s) \ln \frac{s}{r} ,
  \end{equation}
  with $\tilde{C}$ a numerical constant given by Eq.~(\ref{eq:tildeC}). Again
  Eq.~(\ref{eq:sol3}) closely resembles Eqs.~(\ref{eq:sol1}) and
  (\ref{eq:sol2}).
\end{itemize}

The physical reason for the occurrence of the singularity in the limit
$\lambda,\; L,\; R_{\rm drop} \rightarrow +\infty$ is that a ``restoring
force'' far from the particle is required to yield a well--defined
unperturbed interface which allows one to determine the
deformation $u(r)$ unambiguously. For example, if one takes the limit
$\lambda \rightarrow +\infty$ in the Young--Laplace
Eq.~(\ref{eq:younglaplace}), it is inconsistent to impose the boundary
condition $u(r\rightarrow +\infty)=0$ in the corresponding solution.
The special case $\varepsilon_\Pi=\varepsilon_F$, however, is not
singular; this will be discussed in Sec.~\ref{sec:discussion}.  Furthermore, the
comparison of Eqs.~(\ref{eq:sol1}), (\ref{eq:sol2}), and
(\ref{eq:sol3}) demonstrates that the functional form of the
intermediate asymptotic behavior is independent of how the restoring
force is implemented.
%
This corresponds to the so--called intermediate asymptotic behavior of
the second kind \cite{Barenblatt}, characterized by the following
features:
\begin{enumerate}
\item There is a length scale ($\lambda$, $L$, or $R_{\rm drop}$)
  which is much larger than the other length scales of the system
  under consideration and which seems --- at first sight --- to be
  irrelevant.
\item Nevertheless, this length scale determines the dominant
  logarithmic (or more generally, the power--law) dependence.
\item The detailed physical origin of this length scale (in the
  examples we have considered, gravity, pinning of a reference flat or
  curved interface) does not matter.
\end{enumerate}
Well known examples of this kind of asymptotic behavior are critical
phenomena in phase transitions \cite{Goldenfeld}. In that case, it is
the microscopic length scale given by the amplitude of the
  correlation length which cannot be set to zero although it is
much smaller than the correlation length
itself. This microscopic length scale is required to formulate the
power--law behavior of certain properties of the system, but its
detailed physical origin is unimportant for the universal decay exponents of the
power laws.

\section{
        Effective interaction potential of two floating colloids}
\label{sec:2coll}

\begin{figure}
  \begin{center}
    \epsfig{file=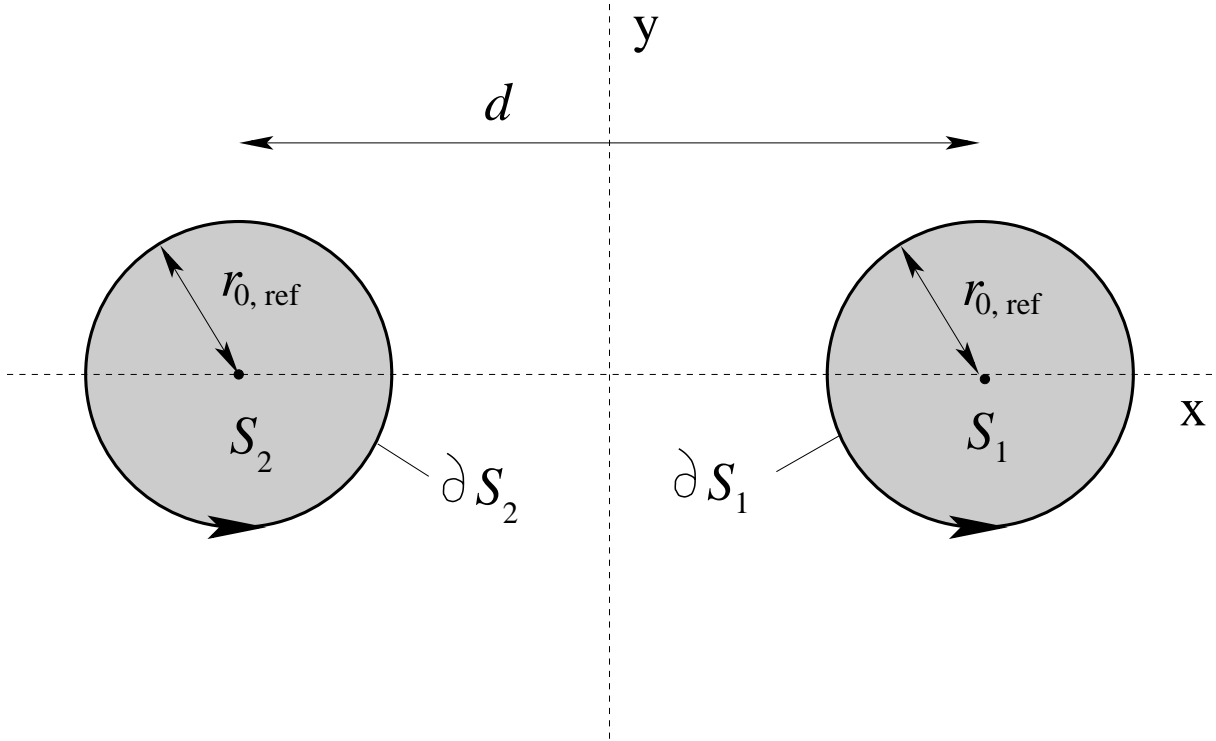,width=.9\textwidth}
    \caption{
      Top view (projection onto the plane $z=0$) of the reference
      configuration with two colloids. $d$ is the distance between the
      colloid centers. $S_1$ and $S_2$ are disks of radius $r_{0, \rm
        ref}$, the corresponding circumferences (counterclockwise) are
      $\partial S_1$ and $\partial S_2$. The projection of the interface
      is $S_{\rm men, ref} = \mathbb{R}^2\backslash(S_1 \bigcup S_2)$.}
    \label{fig:2coll_top}
  \end{center}
\end{figure}

In this section we consider the equilibrium state of two identical
colloids floating at the interface at a fixed lateral distance $d$ and
compute the effective interaction potential $V_{\rm men}(d)$ generated
by the meniscus.  The free energy can be derived along the same lines
leading to Eq.~(\ref{eq:F}), but with due account for the fact that in
the presence of two colloids the meniscus slope no longer exhibits
rotational symmetry. Nonetheless $V_{\rm men}$ depends only on the
distance $d$ between the centers of the two spheres. The reference
configuration is that of two colloids floating on a planar
interface with the corresponding reference free energy being independent of $d$. 
In this case one has for the free energy relative to that
of the reference configuration a contribution ${\cal F}_{\rm men}+{\cal
F}_{\rm vol}+{\cal F}_{\rm inter}$ from the meniscus and a
contribution of the form ${\cal F}_{\rm cont}+{\cal F}_{\rm coll}$
from each colloid.
(The total free energy includes also the direct interaction between the colloids;
this contribution will be considered in Subsec.~\ref{sec:elec}.)
 From Eqs.~(\ref{eq:Fcont_aniso}), (\ref{eq:Fmen}),
(\ref{eq:Fvol}), (\ref{eq:Finter}), and (\ref{eq:Fcoll}) one obtains:
\begin{equation}
  \label{eq:F2}
  {\cal \hat{F}} \simeq 
  \gamma \int_{\mathbb{R}^2\backslash(S_1 \bigcup S_2)} \!\!\!\!\!\! dA \; 
  \left[ \frac{1}{2} |\nabla \hat{u}|^2 + \frac{\hat{u}^2}{2 \lambda^2} - 
    \frac{1}{\gamma} \hat{\Pi} \, \hat{u} \right] + \sum_{\alpha=1,2} \left\{
    \frac{\gamma}{2 r_{0, \rm ref}} \oint_{\partial S_\alpha} \!\!\! d\ell \; [\Delta \hat{h}_\alpha - \hat{u}]^2 
    - \hat{F} \Delta \hat{h}_\alpha \right\} .
\end{equation}
Here, $\hat{u}$ is the meniscus profile in the presence of two
colloids, $\Delta \hat{h}_\alpha$ are the corresponding heights,
$\hat{\Pi}$ is the vertical force per unit area on the meniscus in the
reference configuration, and $\hat{F}_\alpha$ is the force on each
colloid. By symmetry, one has $\Delta \hat{h}_1=\Delta \hat{h}_2$ and
$\hat{F}_1=\hat{F}_2$. $S_\alpha$ are the circular disks delimited by
the contact lines of the colloids in the reference configuration,
$\partial S_\alpha$ are the contact lines, with the convention that we
trace them counterclockwise, and $S_{\rm men, ref} =
\mathbb{R}^2\backslash(S_1 \bigcup S_2)$ (see Fig.~\ref{fig:2coll_top}).

\subsection{Minimization of the free energy within the superposition approximation}

The equilibrium configuration is the minimum of the free
energy given by Eq.~(\ref{eq:F2}). The minimization procedure follows closely
Subsec.~\ref{sec:minF}.  First, minimizing with respect to $\Delta
\hat{h}_\alpha$ at fixed meniscus height $\hat{u}$ leads to the height
of the colloids,
\begin{equation}
  \label{eq:h_eq2}
  \frac{\partial {\cal F}}{\partial (\Delta h_\alpha)} = 0 \qquad \Rightarrow \qquad 
  \Delta \hat{h}_\alpha = \bar{\hat{u}}_\alpha - \varepsilon_{\hat{F}}\; r_{0, \rm ref} ,
\end{equation}
where
\begin{equation}
  \bar{\hat{u}}_\alpha := \frac{1}{2 \pi r_{0, \rm ref}} 
  \oint_{\partial S_\alpha} \!\!\! d\ell \; \hat{u}
\end{equation}
is the mean height of the contact line. 
In the next step, we minimize with respect to $\hat{u}$ at fixed
$\Delta \hat{h}_\alpha$.  Variation in the interior of the domain
$\mathbb{R}^2\backslash(S_1 \bigcup S_2)$ provides a second--order
partial differential equation,
\begin{equation}
  \label{eq:younglaplace2}
  \nabla^2 \hat{u} - \frac{1}{\lambda^2} \hat{u} = - \frac{1}{\gamma} \hat{\Pi} ,
\end{equation}
while variation at the boundary $\partial S_1 \bigcup \partial S_2$ provides the
following {\it transversality conditions} \cite{CoHi}:
\begin{equation}
  \label{eq:trans}
  \frac{\partial \hat{u} ({\bf r})}{\partial n_\alpha} = 
  \frac{\hat{u}({\bf r})-\Delta \hat{h}_\alpha}{r_{0, \rm ref}} =
  \varepsilon_{\hat{F}} + 
  \frac{\hat{u}({\bf r})-\bar{\hat{u}}_\alpha}{r_{0, \rm ref}} , 
        \qquad {\bf r}=(x,y) \in \partial S_\alpha ,
\end{equation}
where the last equality follows from Eq.~(\ref{eq:h_eq2}). In this
expression, $\partial/\partial n_\alpha$ is the derivative in the
outward normal direction of $\partial S_\alpha$. (In this way, the
triad (${\bf e}_n, {\bf e}_t, {\bf e}_z$) is right-handed, where ${\bf
e}_n$ is the unit vector in the outward normal direction, ${\bf e}_t$
is the unit vector in the counterclockwise tangent direction, and
${\bf e}_z$ is the unit vector in the positive $z$--direction.) When
applying the transversality condition, in the context of Gauss'
theorem one must keep in mind that the boundary of the region
$\mathbb{R}^2\backslash(S_1 \bigcup S_2)$ consists of the contours
$\partial S_\alpha$ traced in {\em clockwise direction} with the
normals directed towards the interior of $S_\alpha$. (The boundary at
$r \rightarrow \infty$ does not contribute due to
Eq.~(\ref{eq:u_infty2}).) In the special case of a single colloid,
rotational invariance reduces Eq.~(\ref{eq:trans}) to
Eq.~(\ref{eq:u'0}). Finally, one has the additional boundary condition
\begin{equation}
  \label{eq:u_infty2}
  \lim_{r \to \infty} \hat{u}({\bf r}) = 0 . 
\end{equation}
The solution of Eq.~(\ref{eq:younglaplace2}) with the boundary
conditions given by Eqs.~(\ref{eq:trans}) and (\ref{eq:u_infty2}) is a difficult 
task. (We are only aware of the --- already very involved --- solution of the homogeneous
Eq.~(\ref{eq:younglaplace2}), i.e., $\hat \Pi =0$, with the simplified
boundary condition $\partial \hat u/\partial n_\alpha = \varepsilon_{\hat F}$
\cite{Kra91}.)
For the present purpose, one can use the so-called {\em superposition
  approximation} \cite{Nico49,CHW81,PKDN93}, which yields the correct
solution in the asymptotic limit of large separation $d \gg R$ between the
colloids. Let ${u}_\alpha$ denote the equilibrium
meniscus profile as if colloid $\alpha$ was alone, 
with ${\Pi}_\alpha$ and ${F}_\alpha$ denoting the
corresponding forces. The superposition approximation then reads:
\begin{eqnarray}
  \label{eq:superposition}
  \hat{u} & \simeq & {u}_1 + {u}_2 , \nonumber \\
  \hat{\Pi} & \simeq & {\Pi}_1 + {\Pi}_2 , \\
  \hat{F} & \simeq & {F}_1 = {F}_2 . \nonumber
\end{eqnarray}
Notice that the fields $u_\alpha({\bf r})$ and $\Pi_\alpha({\bf r})$
are defined in the domain $\mathbb{R}^2\backslash S_\alpha$, while the
fields $\hat{u}({\bf r})$ and $\hat{\Pi}({\bf r})$ are
defined 
in the smaller domain 
$\mathbb{R}^2\backslash(S_1 \bigcup S_2)$.
Equations~(\ref{eq:younglaplace2}) and (\ref{eq:u_infty2}) are fulfilled by
this approximate solution, but the boundary
condition~(\ref{eq:trans}) is violated: using
Eqs.~(\ref{eq:superposition}) and the boundary
condition in Eq.~(\ref{eq:u'0}) for the single--colloid solution, one obtains
\begin{equation}
  \label{eq:trans1}
  \left[ \frac{\partial \hat{u} ({\bf r})}{\partial n_1} - \varepsilon_{\hat{F}} - 
    \frac{\hat{u}({\bf r})-\bar{\hat{u}}_1}{r_{0, \rm ref}} \right]_{{\bf r} \in \partial S_1} 
  \simeq \left[ \frac{\partial {u}_2 ({\bf r})}{\partial n_1} - 
    \frac{1}{r_{0, \rm ref}} \left\{ {u}_2 ({\bf r})-
      \frac{1}{2 \pi r_{0, \rm ref}} \oint_{\partial S_1} \!\!\! d\ell \; {u}_2 \right\} 
  \right]_{{\bf r} \in \partial S_1} ,
\end{equation}
and a similar expression for the other colloid with the indices 1 and
2 interchanged. In general, this expression does not vanish as required by 
Eq.~(\ref{eq:trans}).
If $d$ is large, Eq.~(\ref{eq:trans1}) can be evaluated by expanding ${u}_2$
into a Taylor series around the center of colloid 1, yielding to lowest
order (${\bf e}_1$ is the outward directed normal unit vector of
$\partial S_1$)
\begin{equation}
  \label{eq:correction_sup}
  \left[ \frac{\partial \hat{u} ({\bf r})}{\partial n_1} - \varepsilon_{\hat{F}} - 
    \frac{\hat{u}({\bf r})-\bar{\hat{u}}_1}{r_{0, \rm ref}} \right]_{{\bf r} \in \partial S_1} 
  \simeq \frac{1}{4} r_{0, \rm ref} [ {\bf e}_1 {\bf e}_1 : \nabla \nabla {u}_2 (d) + 
  \nabla^2 {u}_2 (d) ] 
\end{equation}
in dyadic notation. Inserting the single--colloid
solution given in Eq.~(\ref{eq:sol1}), one finds that this expression decays like
$d^{-2}$ if $\varepsilon_F \neq \varepsilon_\Pi$, and like
${\Pi}(d) \sim d^{-n}$ if $\varepsilon_F = \varepsilon_\Pi$.

The superposition solution can be used to determine the vertical
displacement according to Eq.~(\ref{eq:h_eq2}): 
\begin{equation}
  \label{eq:z_superposition}
  \Delta \hat{h}_\alpha \simeq \Delta {h} + \bar{{u}},
\end{equation}
where $\Delta {h}$ is the relative vertical displacement of an
isolated colloid (Eq.~(\ref{eq:h_eq})) and
\begin{equation}
  \label{eq:meanu}
  \bar{{u}} := \frac{1}{2 \pi r_{0, \rm ref}} \oint_{\partial S_1} \!\!\! d\ell \; {u}_2
\end{equation}
is the average of the single--colloid meniscus height at the contact
line of the other colloid. 

\subsection{Effective interaction potential}
\label{sec:Vmen}

The meniscus--induced effective potential between the two colloids
(without their direct interaction) is defined as
\begin{equation}
  \label{eq:Veff_a}
  V_{\rm men}(d) := {\cal \hat{F}} - {\cal F}_1 - {\cal F}_2 ,
\end{equation}
where ${\cal \hat{F}}$ is the free energy of the two--colloid
equilibrium configuration (Eq.~(\ref{eq:F2})) while ${\cal F}_1 =
{\cal F}_2$ is the single--colloid equilibrium free energy
(Eq.~(\ref{eq:F})). As noted before the energy of the reference
configuration is independent of the separation $d$ and drops out from 
Eq.~(\ref{eq:Veff_a}).
We insert
Eqs.~(\ref{eq:superposition}), (\ref{eq:z_superposition}), and  (\ref{eq:meanu})
into Eq.~(\ref{eq:Veff_a})
and exploit the invariance of the free energy under exchange of the colloids, i.e., the
symmetry under exchanging indices $1 \leftrightarrow 2$. After carrying out some
analytic manipulations
one finds the following expression for the effective potential:
\begin{eqnarray}
  \label{eq:Vmen1}
  V_{\rm men} (d) & \simeq  &  
  \int_{\mathbb{R}^2\backslash(S_1 \bigcup S_2)} \!\!\!\!\!\! dA \; \left[ 
    \gamma (\nabla {u}_1) \cdot (\nabla {u}_2) + 
    \frac{\gamma}{\lambda^2} {u}_1 {u}_2 - 
    2 {\Pi}_1 {u}_2 \right] \nonumber \\
  & & \mbox{} - \int_{S_1} \!\!\! dA \; \left[ 
    \gamma |\nabla {u}_2|^2 + 
    \frac{\gamma}{\lambda^2} {u}_2^2 - 
    2 {\Pi}_2 {u}_2 \right] \nonumber \\
  & & \mbox{}  + 
  \frac{\gamma}{r_{0, \rm ref}} \oint_{\partial S_1} \!\!\! d\ell \; [\bar{{u}} - {u}_2]^2 - 2 F \bar{{u}}\;.
\end{eqnarray}
The first integral accounts for the change in surface energy, gravitational
potential energy, and surface--stress potential energy of the meniscus
due to the overlap of the meniscus deformations caused by the two colloids.
The second integral is the corresponding change due to the fact that
the interface is reduced by an amount $S_1$ compared to the
single--colloid case because of the presence of the second colloid.
The third integral is the change in
surface free energy of one colloid due to the extra meniscus deformation
induced by the second colloid.
The last term is the change in energy due to the vertical
displacement of one colloid by this extra meniscus deformation. 

For the mathematical manipulations to follow, it is suitable to
rewrite Eq.~(\ref{eq:Vmen1}) by applying Gauss' theorem to the integrals
involving $\nabla {u}$ and by using the fact that the functions $u_\alpha$ fulfill
Eq.~(\ref{eq:younglaplace}) individually, e.g.,
\begin{eqnarray}
  \gamma \int_{\mathbb{R}^2\backslash(S_1 \bigcup S_2)} \!\!\!\!\!\! dA \;
  (\nabla {u}_1) \cdot (\nabla {u}_2) & = &
  \gamma \int_{\mathbb{R}^2\backslash(S_1 \bigcup S_2)} \!\!\!\!\!\! dA \;
  [ \nabla \cdot ({u}_2 \nabla {u}_1) - {u}_2 \nabla^2 {u}_1 ] = \nonumber \\
  & & \mbox{} - \gamma \oint_{\partial S_1} \!\!\! d\ell \; \frac{\partial ({u}_1 {u}_2)}{\partial n_1} +
  \int_{\mathbb{R}^2\backslash(S_1 \bigcup S_2)} \!\!\!\!\!\! dA \; \left[ {\Pi}_1 {u}_2 - 
    \frac{\gamma}{\lambda^2}  {u}_1 {u}_2 \right] . \nonumber \\
  \end{eqnarray}
Thus one obtains from Eq.~(\ref{eq:Vmen1})
\begin{eqnarray}
  \label{eq:Vmen2}
  V_{\rm men} (d) & \simeq  &  
  \mbox{} - \int_{\mathbb{R}^2\backslash(S_1 \bigcup S_2)} \!\!\!\!\!\! dA \; {\Pi}_1 \, {u}_2 
  + \int_{S_1} \!\!\! dA \; {\Pi}_2 \, {u}_2 \nonumber \\
  & & \mbox{} - \gamma \oint_{\partial S_1} \!\!\! d\ell \; \frac{\partial ({u}_1 {u}_2)}{\partial n_1}
  - \frac{1}{2} \gamma \oint_{\partial S_1} \!\!\! d\ell \; \frac{\partial {u}_2^2}{\partial n_1} \nonumber \\
  & & \mbox{} + 
  \frac{\gamma}{r_{0, \rm ref}} \oint_{\partial S_1} \!\!\! d\ell \; [\bar{{u}} - {u}_2]^2 
  - 2 F \bar{{u}} .
\end{eqnarray}
In this form all the integrals, except the first one, are performed over
bounded domains. This allows one to carry out a
Taylor expansion which yields a uniformly valid asymptotic
expansions of the terms as $d \to \infty$.
In the following calculations, the integrals are evaluated in polar
coordinates with the origin at the center of $S_\alpha$ and ${\bf
r}_\alpha$ is the position vector with respect to this origin. In particular, ${\bf
r}_2 = d {\bf e}_x$ is the center of colloid 1 with respect to the
center of colloid 2, and ${\bf r}_1 = - d {\bf e}_x$ is the center of
colloid 2 with respect to the center of colloid 1
(Fig.~\ref{fig:2coll_top}) so that, e.g., $u_\alpha = u(|{\bf
r}_\alpha|)$.
\begin{enumerate}
\item In order to compute the leading asymptotic behavior of the first term
in Eq.~(\ref{eq:Vmen2})  as $d \to \infty$ we distinguish two cases:
\begin{enumerate} 
\item $\varepsilon_\Pi \neq \varepsilon_F$: In this case, ${u} \sim
\log{r}$ (Eq.~(\ref{eq:sol1})) while ${\Pi} \sim r^{-n}$ ($n>2$) as $r
\to \infty$.  Asymptotically the main contribution stems 
from the region near $S_1$ and
\begin{equation}
  \label{eq:pi1u2_a} 
  \int_{\mathbb{R}^2\backslash(S_1 \bigcup S_2)}
  \!\!\!\!\!\! dA \; {\Pi}_1 \, {u}_2 \simeq {u}(d)
  \int_{\mathbb{R}^2\backslash S_1} \!\!\!\!\!\! dA \; {\Pi}_1 = 2 \pi
  \gamma r_{0, \rm ref}\; \varepsilon_\Pi\; {u}(d) , 
\end{equation}
  employing the definition~(\ref{eq:epsPi}). It will turn out that there is
  no need to compute also the next--to--leading term in order to obtain
  $V_{\rm men} (d \to \infty)$ as the leading term will be 
  part of the leading contribution to $V_{\rm men}$ and it will not be cancelled
  by other contributions.
    
  \item $\varepsilon_\Pi = \varepsilon_F$: In this case ${u} \sim r^{2-n}$
    if ${\Pi}(r \to \infty) \sim r^{-n}$.  The regions
    contributing mainly to the integral are the neighborhoods of $S_1$
    and $S_2$. We employ the Taylor expansion of the integrand (which
    is valid up to a maximum order depending on how fast $\Pi$ decays)
    to evaluate the leading and the next--to--leading contributions
    if $n>4$ (using dyadic notation):
    \begin{eqnarray}
      \int_{\mathbb{R}^2\backslash(S_1 \bigcup S_2)} \!\!\!\!\!\! dA \; {\Pi}_1 \, {u}_2 & \simeq  &
      \int_{\mathbb{R}^2\backslash S_1} \!\!\!\!\!\! dA \; {\Pi}_1 \left[{u}_2 +
      {\bf r}_1 \cdot (\nabla {u})_2 +
      \frac{1}{2} {\bf r}_1 {\bf r}_1 : (\nabla \nabla {u})_2 + 
      \dots \right]_{{\bf r}_2 = d {\bf e}_x} \nonumber \\
    & & \mbox{} + \int_{\mathbb{R}^2\backslash S_2} \!\!\!\!\!\! dA \; {u}_2 \left[{\Pi}_1 +
      \dots \right]_{{\bf r}_1 = - d {\bf e}_x} \nonumber \\
    & \simeq & 2 \pi \gamma r_{0, \rm ref}\; \varepsilon_\Pi \;{u}(d) +
    \frac{1}{2} \pi \nabla^2 {u} (d) \int_{r_{0,\rm ref}}^\infty dr\; r^3\,\Pi(r) \nonumber \\
    & & \mbox{} + {\Pi}(d) \int_{\mathbb{R}^2\backslash S_2} \!\!\!\!\!\! dA \; {u}_2 + \dots \; .
    \end{eqnarray}
    We can simplify the result further by evaluating the last integral
    by repeated partial integration
    with the explicit solution for ${u}$ given in Eq.~(\ref{eq:sol1}): 
    \begin{equation}
      \label{eq:int_uda}
      \int_{\mathbb{R}^2\backslash S_2} \!\!\!\!\!\! dA \; {u} (r_2) = \frac{1}{2} \pi r_{0, \rm ref}^3 \left[
        \varepsilon_\Pi - \frac{2 {u}_0}{r_{0,\rm ref}} - \frac{1}{\gamma r_{0,\rm ref}^3} 
        \int_{r_{0,\rm ref}}^\infty dr\; r^3\,\Pi(r) \right] .
    \end{equation}
    One finally obtains
    \begin{eqnarray}
      \label{eq:pi1u2_b}
      \int_{\mathbb{R}^2\backslash(S_1 \bigcup S_2)} \!\!\!\!\!\! dA \; {\Pi}_1 \, {u}_2 & \simeq &
      2 \pi \gamma r_{0, \rm ref}\; \varepsilon_\Pi\; {u}(d) \\
      & & \mbox{} + \left[ 2 \int_{\mathbb{R}^2\backslash S_2} \!\!\!\!\!\! dA \; {u} (r_2) - 
        \frac{1}{2} \pi r_{0, \rm ref}^3 \left( \varepsilon_\Pi - \frac{2 {u}_0}{r_{0,\rm ref}} \right) 
      \right] {\Pi} (d) + \dots \nonumber
    \end{eqnarray}
    where we have employed Eq.~(\ref{eq:younglaplace2}) (with
    $\lambda^{-1} = 0$ for simplicity). Convergence of the integral of
    ${u}$ imposes the more stringent constraint $n>4$ on the decay
    of ${\Pi}$ (see Eq.~(\ref{eq:int_uda})).
  \end{enumerate}
  The validity of formulae~(\ref{eq:pi1u2_a}) and (\ref{eq:pi1u2_b})
  has been checked explicitly by comparing their predictions with the numerical
  evaluation of the integral for a surface stress of the form $\Pi
  \propto r^{-n}$ with $n>4$.
  
\item The second term in Eq.~(\ref{eq:Vmen2})
  can be estimated in the limit $d \to \infty$ by expanding the
  integrand into a Taylor series, which is uniformly valid in the
  integration domain:
  \begin{eqnarray}
    \label{eq:pi2u2}
    \int_{S_1} \!\!\! dA \; {\Pi}_2 \, {u}_2 & \simeq &
    \int_{S_1} \!\!\! dA \; \left[ {\Pi}_2 \, {u}_2 + 
      \dots \right]_{{\bf r}_2 = d {\bf e}_x} 
    \simeq \pi r_{0, \rm ref}^2 \, {\Pi} (d) \, {u} (d) + 
      \dots \; .
  \end{eqnarray}
  
\item The third term in Eq.~(\ref{eq:Vmen2}) reads:
  \begin{eqnarray}
    \label{eq:u1u2}
    \oint_{\partial S_1} \!\!\! d\ell \; \frac{\partial ({u}_1 {u}_2)}{\partial n_1} & = &
    \int_0^{2 \pi} \!\!\! d\varphi \; 
    \left[ {\bf r}_1 \cdot \nabla ({u}_1 {u}_2) \right]_{r_1=r_{0, \rm ref}} \\
    & \simeq & \int_0^{2 \pi} \!\!\! d\varphi \; \left\{ ({\bf r}_1 \cdot \nabla) 
    \left[ {u}_1 \left( 
   \phantom{ \frac{1}{2} } \!\!\!
   {u}_2 + {\bf r}_1 \cdot (\nabla {u})_2 + 
   \right.\right.\right.\nonumber \\
    & & \qquad \qquad \left. \left. \left.
      \frac{1}{2} {\bf r}_1 {\bf r}_1 : (\nabla \nabla {u})_2
      + \dots \right)_{{\bf r}_2=d {\bf e}_x} \right] \right\}_{r_1=r_{0, \rm ref}} \nonumber \\
  & \simeq & 2 \pi r_{0, \rm ref}  {u}(d) \left. \frac{d {u}}{d r} \right|_{r=r_{0, \rm ref}} + 
  \frac{\pi}{2} r_{0, \rm ref} \nabla^2 {u}(d) \left. \frac{d (r^2 {u})}{d r} \right|_{r=r_{0, \rm ref}} + 
  \dots \nonumber \\
  & \simeq & 2 \pi r_{0, \rm ref} \varepsilon_F {u}(d) - 
  \frac{\pi}{2\gamma} r_{0, \rm ref}^3 \left(\varepsilon_F + \frac{2{u}_0}{r_{0, \rm ref}} \right) {\Pi}(d) + \dots
  \end{eqnarray}
  using the boundary condition for ${u}$ in Eq.~(\ref{eq:u'0}).

\item The fourth term in Eq.~(\ref{eq:Vmen2}) is evaluated analogously:
  \begin{eqnarray}
    \oint_{\partial S_1} \!\!\! d\ell \; \frac{\partial {u}_2^2}{\partial n_1} & = &
    \int_0^{2 \pi} \!\!\! d\varphi \; 
    \left[ {\bf r}_1 \cdot \nabla {u}_2^2 \right]_{r_1=r_{0, \rm ref}} \\
    & \simeq & \int_0^{2 \pi} \!\!\! d\varphi \; 
    \left\{ {\bf r}_1 \cdot \left[ (\nabla {u}^2)_2 + {\bf r}_1 \cdot (\nabla \nabla {u}^2)_2
      + \dots \right]_{{\bf r}_2=d {\bf e}_x} \right\}_{r_1=r_{0, \rm ref}} \nonumber \\
    & \simeq & \pi r_{0, \rm ref}^2 \nabla^2  {u}^2 (d) + \dots \; .
  \end{eqnarray}

\item The fifth term in Eq.~(\ref{eq:Vmen2}) is given by
  \begin{eqnarray}
    \label{eq:varu_taylor}
    \frac{1}{r_{0,\rm ref}} \oint_{\partial S_1} \!\!\! d\ell \; [{u}_2 - \bar{{u}}]^2 & = &
    \int_0^{2 \pi} \!\!\! d\varphi \; {u}_2^2 - 2 \pi \bar{{u}}^2 \nonumber \\
    & \simeq & \frac{1}{2} \pi r_{0, \rm ref}^2 \left[ \nabla^2 {u}^2 (d) - 
     2 {u} \nabla^2 {u} (d) \right] + \dots \nonumber \\
   & \simeq & \frac{1}{2} \pi r_{0, \rm ref}^2 \left[ \nabla^2 {u}^2 (d) + 
     \frac{2}{\gamma} {\Pi} (d) \, {u} (d) \right] + \dots  \;.
  \end{eqnarray}

\item In order to evaluate the sixth term in Eq.~(\ref{eq:Vmen2}), we expand $u_2$ in the
  definition given by Eq.~(\ref{eq:meanu}):
  \begin{equation}
    \label{eq:meanu_taylor}
    \bar{{u}} = \frac{1}{2 \pi r_{0, \rm ref}} \oint_{\partial S_1} \!\!\! d\ell \; {u}_2 \simeq
    {u}(d) + \frac{1}{4} r_{0, \rm ref}^2 \nabla^2 {u} (d) + \dots 
    \simeq {u}(d) - \frac{1}{4\gamma} r_{0, \rm ref}^2 {\Pi} (d) + \dots  \; .
  \end{equation}
\end{enumerate}
The asymptotic behavior of the effective potential is finally obtained
from Eq.~(\ref{eq:Vmen2}) by collecting all terms. There are two
qualitatively different cases:
\begin{itemize}
\item \underline{$\varepsilon_\Pi \neq \varepsilon_F$}: The limit $\lambda \to
  \infty$ is singular and the single--colloid meniscus profile
  exhibits a logarithmic dependence. One finds that the leading 
  contribution is provided by the terms proportional to ${u}(d)$:
  \begin{equation}
    \label{eq:logV}
    V_{\rm men} (d) \simeq  2 \pi \gamma r_{0, \rm ref}\; (\varepsilon_F - \varepsilon_\Pi)
     \; {u}(d) 
    \simeq \mbox{} - 2 \pi \gamma r_{0, \rm ref}^2\; (\varepsilon_F - \varepsilon_\Pi)^2\; 
    \ln\frac{C \lambda}{d}\;, \qquad (R \ll d \ll \lambda) ,
  \end{equation}
  which represents a long-ranged attractive effective potential,
  irrespective of the sign of the forces $F$ and $\Pi$ acting on the
  system.
  
\item \underline{$\varepsilon_\Pi = \varepsilon_F$}: The
  single--colloid meniscus profile decays like ${u}(d) \sim
  d^{2-n}$ if ${\Pi}(d) \sim d^{-n}$. The leading contribution
  is proportional to ${\Pi}(d)$, because the terms proportional to
  ${u}(d)$ cancel each other:
  \begin{equation}
    \label{eq:shortrangeV}
    V_{\rm men} (d) \simeq  \mbox{} - 2 \, {\Pi}(d)
\int_{\mathbb{R}^2\backslash S} \!\!\!\!\!\! dA \; {u} , \qquad (R \ll d \ll \lambda) .
  \end{equation}
  This correspond to a shorter--ranged effective interaction, which in
  principle can be either attractive or repulsive depending on the
  precise form of the function ${\Pi}(r)$.
  In the particular case that ${\Pi}(r)$ decays monotonically to
  zero, e.g., ${\Pi}(r) \propto r^{-n}$, it is easy to check that
  $V_{\rm men}$ amounts to a repulsive force, because the sign of
  ${u}$ is opposite to the sign of ${\Pi}$.
  
  We have seen that the error committed by the superposition
  approximation in satisfying one of the boundary conditions, Eq.~(\ref{eq:correction_sup}), decays like
  ${\Pi}(d)$, too. This suggests that the corrections to the
  superposition approximation could modify the precise value of the
  constant factor acting as an amplitude in Eq.~(\ref{eq:shortrangeV}), 
and {\it a priori} it
  cannot be excluded that there are cancellations leading to a
  vanishing amplitude, and therefore to an even faster decay for large $d$. Thus
  the superposition approximation might
  not be reliable enough for calculating $V_{\rm men}$ if
  $\varepsilon_\Pi = \varepsilon_F$.
\end{itemize}

If one traces back the origin of the dominant contributions to $V_{\rm
  men}$, one finds that in both cases only the first, third, and sixth
term in Eq.~(\ref{eq:Vmen2}) are relevant. They correspond physically
to the effect of the overlap of the two single--colloid meniscus
profiles and the change of the colloid height (first integral and the
term $- 2 F \bar{{u}}$ in Eq.~(\ref{eq:Vmen1}), respectively).


\section{Applications and discussion}
\label{sec:discussion}

Equations (\ref{eq:logV}) and  (\ref{eq:shortrangeV}) describe the asymptotic
behavior of the meniscus--induced
effective intercolloidal potential and thus represent a central result of our
analysis. They provide the explicit functional dependence on  an arbitrary
stress field $\Pi(r)$ which decays sufficiently fast. The assumptions
entering their derivation are (i) that the deviations of the meniscus profile 
from the reference configuration are small, allowing one to confine the analysis to 
a free energy expression which is
quadratic in the deviations, and (ii) the superposition approximation,
which expresses the two--colloid equilibrium state in terms of the
single--colloid state. 
The analysis shows  that the limit $\lambda \to +\infty$ is non--singular only 
in the case $\varepsilon_F = \varepsilon_\Pi$.

\subsection{Flotation force}

Equation (\ref{eq:logV}) can be used to determine the
flotation force. There are no stresses acting at the meniscus, $\Pi
\equiv 0$, while the force $F$ on the colloids is due to their weight
corrected by the buoyancy force. 
Accordingly, Eq.~(\ref{eq:logV}) reduces to the flotation potential
\begin{eqnarray}
  \label{eq:flotV}
  V_{\rm flot}(d) &=& 2 \pi \gamma r_{0,\rm ref} \varepsilon_F \, u(d)\; \\
 & =&
  2 \pi \gamma Q^2 \ln\frac{d}{C \lambda}\;, \qquad (R\ll d \ll \lambda) , \nonumber
\end{eqnarray}
where $Q := \varepsilon_F r_{0, \rm ref}$ is known as the capillary
charge \cite{Kra00}, by analogy with two--dimensional
electrostatics. 
The asymptotic form for $d \ll \lambda$ originates from a potential
proportional to the modified Bessel function
$K_0(d/\lambda)$ (see Eq.~(\ref{eq:solBessel})) \cite{Nico49,CHW81}.
The order of magnitude of the capillary charge is $Q \sim r_{0, \rm ref} (R/\lambda)^2$.
For a typical value \mbox{$\gamma = 0.05$ N m$^{-1}$} at room
temperature and for colloids with a mass density of the order of
\mbox{$1$ g cm$^{-3}$}, the prefactor of the logarithm can be estimated as
\begin{equation}
  2 \pi \gamma Q^2 \sim \left(\frac{R}{10 \rm \mu m}\right)^6 k_B T .
\end{equation}
Therefore, compared with the thermal energy, the flotation force is negligible 
for submicrometer--sized colloids.

\subsection{Electrically charged colloids}
\label{sec:elec}

\begin{figure}
  \begin{center}
    \epsfig{file=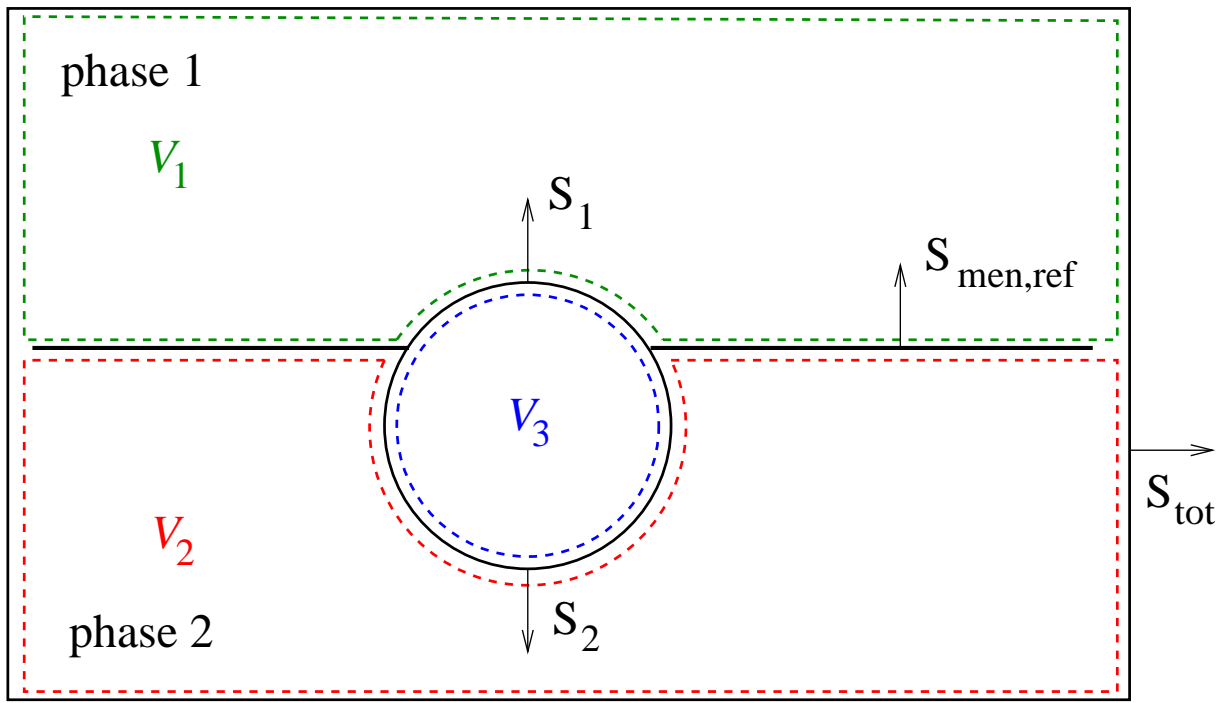,width=.9\textwidth}
    \caption{ (color online)
      In the reference configuration the whole system is divided into
      volumes $V_1$, $V_2$, and $V_3$. Volume $V_1$ (enclosed by the
      upper dashed curve) includes phase 1, volume $V_2$ (enclosed by
      the lower dashed curve) includes phase 2, and volume $V_3$
      includes the interior of the colloid. The arrows indicate the
      direction in which the surfaces (including the infinitesimally
      displaced ones) are oriented: $S_{\rm tot}$
      encloses the whole system, $S_{\rm men, ref}$ is the interface
      between phase 1 and phase 2, and $S_{1(2)}$ denotes the interface
      between the colloid and phase 1(2). 
    }
    \label{fig:forcebalance}
  \end{center}
\end{figure}

Another application is the case of electrically charged colloids. If
one of the liquid phases is water, the charge of the colloid is
screened (the Debye length of pure water is $\approx 1$ $\mu$m and
smaller in the presence of an electrolyte), and the effective electric field is
that of a dipole oriented perpendicularly to the fluid interface
\cite{Hur85,RoEa93,GoHa98,Ave00a,Ave02}. 
The electrostatic field decays as $r^{-3}$ and
the stress on the meniscus as $\Pi(r\to \infty) \propto r^{-6}$. Both the
electrostatic stress and the osmotic ionic pressure decay in the same
manner \cite{For04}. Thus the total intercolloidal potential at intermediate
distances is 
\begin{eqnarray}
  \label{eq:Vtot}
  V_{\rm tot} = a \frac{k_B T}{d^3} + V_{\rm men}\qquad (a>0) \;.
\end{eqnarray} 
If gravity is neglected both as a force on the colloid and as a restoring
force for the interface $(\lambda \to \infty)$, 
then one can indeed show that the ensuing condition
of mechanical isolation (no net force on the system) leads to
$\varepsilon_\Pi = \varepsilon_F$, i.e., precisely the situation for which
the limit $\lambda \to \infty$ is non--singular. To see this, we consider 
 the total stress tensor ${\bf T}$ which
consists of the Maxwell stress tensor (due to the electrostatic field)
and a diagonal osmotic pressure tensor (due to the electrolytes)
\cite{GDLM93,SMN02}.
At interfaces ${\bf T}$ can be discontinuous.
The total volume $V$ of the system is divided into volumes $V_1$,
$V_2$, and $V_3$ (see Fig.~\ref{fig:forcebalance} for the explanation
of the notation in the following equation). The total force reads (the
superscript $^{+(-)}$ denotes evaluation on the positive(negative)
side of the oriented surface, i.e., the side the arrows in Fig.~\ref{fig:forcebalance}
point to (do not point to))
\begin{eqnarray}
\label{eq:forcebalance}
  \oint_{S_{\rm tot}} \!\!\! d{\bf A} \cdot {\bf T} & = &
  \int_{V_1\bigcup V_2\bigcup V_3} \!\!\! dV \; (\nabla \cdot {\bf T}) + 
  \int_{S_{\rm men,ref}\bigcup S_1 \bigcup S_2} \!\!\! d{\bf A} \cdot ({\bf T}^+ - {\bf T}^-) \nonumber \\
  & = &  \int_{V_1\bigcup V_2} \!\!\! dV \; (\nabla \cdot {\bf T}) + 
  \int_{S_{\rm men,ref}} \!\!\! d{\bf A} \cdot ({\bf T}^+ - {\bf T}^-) + \nonumber \\
  & & \left[ \int_{V_3} \!\!\! dV \; (\nabla \cdot {\bf T}) +
    \int_{S_1 \bigcup S_2} \!\!\! d{\bf A} \cdot ({\bf T}^+ - {\bf T}^-) \right] \nonumber \\
  & = & \int_{S_{\rm men, ref}} \!\!\!\!\!\! dA \; \Pi  \, {\bf e}_z  + F {\bf e}_z \; .
\end{eqnarray}
In the first line, we have applied Gauss' theorem with due account of
the possible discontinuities of the tensor ${\bf T}$ accross the
interfaces. In the second line, $\nabla \cdot {\bf
  T} = {\bf 0}$ in the fluid phases $V_1$ and $V_2$,  because the
counterion distribution of the present reference configuration is the
equilibrium distribution and thus locally force free. This distribution is 
considered to be fixed.
The second term in the second line is the
total force on the meniscus (which can have only a normal component),
while the terms in square brackets sum
up to the force $F$ acting on the colloid.

Thus the vertical component of the total force is
\begin{eqnarray}
  {\bf e}_z \cdot \oint_{S_{\rm tot}} \!\!\! d{\bf A} \cdot {\bf T} & = &
  2 \pi \gamma r_{0, \rm ref} (\varepsilon_\Pi - \varepsilon_F) .
  \label{eq:forcebalance1}
\end{eqnarray}
If it vanishes, as it is the case for
an isolated system, then $\varepsilon_\Pi
= \varepsilon_F$. According to Eq.~(\ref{eq:logV}) this implies that the long--ranged
logarithmic contribution to $V_{\rm men}$ is absent and thus 
the limit $\lambda \to \infty$ is regular. Physically this means that there is
no need for a restoring force acting on the fluid--fluid interface
when the deformation is created by localized internal
stresses. Instead, according to Eq.~(\ref{eq:shortrangeV}), one
obtains a potential $V_{\rm men} \propto  d^{-6}$ for the present case of  a dipolar 
electric field (see above). This shorter--ranged potential cannot counterbalance
the direct electrostatic dipolar repulsion
$\sim d^{-3}$. Such a counterbalance would be needed for a straightforward explanation
of the aforementioned experimentally observed attractions.

The line of argument to explain the absence of the logarithm contribution to
$V_{\rm men}$ was already put forward  in Refs.~\cite{Meg03} and \cite{For04}, where
exclusively the case ${\Pi} (r) \propto r^{-6}$ has been addressed. Our
detailed analysis complements and generalizes these contributions. In
Ref.~\cite{Meg03}, $V_{\rm men} (d)$ is estimated 
by taking into account only the degree of freedom ``meniscus profile'',
$u({\bf r })$, and considering only the change in meniscus area due to the
superposition of the dimples. This corresponds 
to retaining only the term $\propto
(\nabla {u}_1) \cdot (\nabla {u}_2)$ in expression~(\ref{eq:Vmen1}). 
Although the type of
$d$--dependence obtained that way is correct, the sign of the force turns out to
be wrong (attraction
instead of repulsion). The reason is that the contributions $\propto \Pi_1
u_2$ and $\propto F \bar{u}$ in Eq.~(\ref{eq:Vmen1}) are equally 
important as the retained term. 
This is taken into account in the more detailed
analysis of Ref.~\cite{For04}, where the limiting case $R = 0$ (point--particle)
is considered from the outset, so that ``height of the
colloid'' is not an independent degree of freedom, i.e., $h =
u(0)$. Correspondingly, $F$ is set to zero ($\varepsilon_F=0$), and
its effect is modelled by a Dirac--delta contribution to the stress
$\Pi(r)$ such that $\varepsilon_\Pi=0$.  The potential $V_{\rm
men}(d)$ calculated this way corresponds to keeping
the three terms mentioned above which are relevant 
(the first and third term in the first integral of Eq.~(\ref{eq:Vmen1})
and the last term of  Eq.~(\ref{eq:Vmen1}))
and to setting $\lambda \to
\infty$, since this limit is regular when
$\varepsilon_F=\varepsilon_\Pi=0$. Our analysis has shown that the terms which
are dropped in the limit $R \to 0$ (second and third integral in
Eq.~(\ref{eq:Vmen1}), second, fourth, and fifth terms in
Eq.~(\ref{eq:Vmen2})) yield indeed a subdominant contribution to
Eq.~(\ref{eq:shortrangeV}). However, the integral appearing as a prefactor in
Eq.~(\ref{eq:shortrangeV}) is divergent for $R \to 0$, so that
in Ref.~\cite{For04} a short--distance cutoff $a$ has been introduced
which is  expected to be of
the order of $R$. The precise value of $a$ depends on the details of the
implementation of this cutoff; for the example used in Ref.~\cite{For04},
the application of Eq.~(\ref{eq:shortrangeV}) yields $a=r_{0, \rm ref}$.

In the presence of gravity mechanical isolation is violated and
as we have shown 
$\varepsilon_\Pi - \varepsilon_F \propto (R/\lambda)^2$. (The
force--balance argument (Eq.~(\ref{eq:forcebalance})) can be easily
generalized to include the effect of the gravitational volume force.)
In the case of a curved reference interface (corresponding to the experiment
reported in Ref.~\cite{Nik02} using colloids trapped at the interface of water droplets
in oil), Eq.~(\ref{eq:forcebalance}) is replaced by
\begin{equation}
  \oint_{S_{\rm tot}} \!\!\! d{\bf A} \cdot {\bf T}  = 
  \int_{S_{\rm men, ref}} \!\!\!\!\!\! dA \; \Pi  \, {\bf e}_r  + F {\bf e}_z \; ,
\end{equation}
where ${\bf e}_r$ is the radial unit vector of the unperturbed
spherical droplet. Force balance in the vertical direction yields a
factor ${\bf e}_r \cdot {\bf e}_z = \cos\psi$ (see
Fig.~\ref{fig:droplet}) in the integral, which can be expanded for
small angles in the limit $R_{\rm drop} \to \infty$.  This leads to 
a curvature--induced correction
$\varepsilon_\Pi - \varepsilon_F \propto (r_{0,\rm ref}/R_{\rm
  drop})^2$ even for mechanical isolation. Thus we see that independently of the details of the
implementation of the distant boundary conditions the logarithmic term
in the potential $V_{\rm men}$ (Eq.~(\ref{eq:logV})) has a strength
which is proportional to (colloid size/system length)$^4$, which is
nevertheless too weak to explain the reported attractive total interaction.

\subsubsection{The experiment reported in Ref.~\cite{Nik02}}

Mechanical isolation is violated in the presence of an external
homogeneous electric field $E {\bf e}_z$. 
For the experimental setup used in  Ref.~\cite{Nik02}
it cannot be ruled out that  such an external field 
may have distorted the measurements \cite{Nik04}. Since this is the
only experiment which provides quantitative information about the
secondary minimum in $V_{\rm tot}$, we discuss the case of an external field 
in more detail. In this experiment position--recording measurements 
were performed on a hexagonal
configuration of seven trapped colloids on a water droplet immersed in oil
$(R_{\rm drop} \approx 32 R=24\mu$m).
The latter was confined between two glass plates and the droplet stuck on
the upper glass plate with a contact angle close to $\pi$. Residual
charges might have resided on the upper plate \cite{Nik04}.  
The measurements yielded the position
of the secondary minimum,  $d_{\rm min} \approx 7.6 R$,
and the curvature at the minimum, $V_{\rm tot}''(d_{\rm min}) \approx
12.94 \, k_B T /R^2$. 
With regard to the latter value we remark that systematic corrections 
can be estimated to lower $V_{\rm tot}''$ by a factor 2 to 3.
These systematic corrections include (i) multiparticle effects and (ii) center--of-mass movement of the droplet. We have estimated the effect of (i) by carrying
out Langevin simulations of the seven particle system using the
intercolloidal potential Eq.~(\ref{eq:Vtot1}) below. As for the effect of (ii),
any shape deformation induced by the moving colloids
changes the center--of--mass position
of the droplet since the droplet is fixed to the 
upper glass plate. The corresponding change in the gravitational energy of the
droplet translates into a weak confining potential for the
colloids which limits the stochastic movement of the center--of--mass 
of the seven colloids.
This effect might be part of an explanation for the absence of 
center--of--mass movement observed in  Ref.~\cite{Nik02}. 
According to Eqs.~(\ref{eq:logV}) and (\ref{eq:Vtot}), 
the total intercolloidal potential in an external field is
\begin{eqnarray}
  \label{eq:Vtot1}
  V_{\rm tot} = a \frac{k_B T}{d^3} - b\ln \frac{C\lambda}{d} \qquad (a,b>0) \;.
\end{eqnarray} 
Using the aforementioned experimental data for $d_{\rm min}$ and 
$V_{\rm tot}''(d_{\rm min})$,
one obtains from Eq.~(\ref{eq:Vtot1}) $b\approx 249\:k_B T$, 
$a \approx 83\: d_{\rm min} ^3$ and $V_{\rm tot}(d_{\rm min}) \approx - 275 \:k_B T$
which is surprisingly deep, even if reduced by a factor 2 to 3 due to the systematic corrections mentioned above.
Furthermore, from Eq.~(\ref{eq:logV})
we deduce $b = 2\pi\gamma r_{0,\rm ref}^2\; (\varepsilon_\Pi - \varepsilon_F)^2$
and with $\gamma\approx 0.05$ N/m we find 
$|\varepsilon_\Pi -\nolinebreak \varepsilon_F| \approx (2 \, \rm{nm}/r_{0,\rm ref})$.
The long--ranged meniscus deformation (see Eq.~(\ref{eq:sol1})) is on the
scale of nanometers. The short--ranged meniscus deformation near the colloid
can only be evaluated with a specific microscopic model for $\Pi(r)$.   
For a rough estimate of $\varepsilon_\Pi$, we consider the colloid charge to be  
concentrated
on the surface. The asymptotic behavior of the  stress tensor in this 
case is given by $\Pi(r) = a/(4\pi)\: (k_B T)/r^6$ \cite{Hur85,For04}.
If we assume this form of the stress tensor to hold for all $r$, we find
$\varepsilon_\Pi \approx 10^{-4}/\sin^5\theta$ for the values of $\gamma$ and
$a$ given above. For not too small contact angles, this {\em a posteriori} justifies 
the perturbative approach which we have adopted.   

 If the system has a net
charge $q$, then $|q E| = 2 \pi \gamma r_{0,\rm ref}
|\varepsilon_\Pi-\varepsilon_F| = \sqrt{2 \pi \gamma  \, b}$ 
(see Eqs.~(\ref{eq:forcebalance1}) and (\ref{eq:logV})). 
Using the values for $b$ and $\gamma$ as given above we find
\begin{equation}
 |q E| \approx 3.6 \times 10^9 \, \rm e V \, m^{-1} .
\end{equation}
For the value $q \approx 2 \times 10^5 \, \rm e$ quoted in
Ref.~\cite{Nik02}, this yields a relatively small electric field $E \approx
1.8 \times 10^4 \, \rm V \, m^{-1}$, indicating how sensitive the
system can be to spurious external fields. Alternatively, an electric
field
$E \sim 10^3 \, \rm V \, m^{-1}$ is sufficient for the
meniscus--induced logarithmic potential to have a depth of the order
$b \sim 1 \, k_B T$. 
Thus the external field offers the possibility to tune easily the capillary 
long--ranged attraction and to manipulate the structures formed by the colloids at
the interface.

\subsubsection{The experiment and analysis reported in Ref.~\cite{Kra04}}

In Ref.~\cite{Kra04}, experimental results for the meniscus deformation around glass
particles of radii 200\dots 300 $\mu$m trapped at water--air or water--oil interfaces 
are reported. The data for the meniscus slope $u'(r_0)$ at the contact circle
imply (Eq.~(\ref{eq:u'0})) 
$\varepsilon_F\approx 0.2$ (water--oil interface), about fifteen times  larger than 
the corresponding $\varepsilon_{F,\rm g}$ caused by gravity alone. 
Furthermore, the reported meniscus shape for one
sample contains a logarithmic part which is consistent with  $\varepsilon_{F}-\varepsilon_\Pi
\approx 0.1 \gg \varepsilon_{F,\rm g}$ (see Eq.~(\ref{eq:sol1})).
As in the experiment analyzed in the previous subsection, the
experimental observations could be understood within the framework of the
theoretical model we have developed in terms of an external electric
field violating mechanical isolation.
%
Yet independently of us, the
authors of Ref.~\cite{Kra04} have developed a theoretical model based on the same physical hypotheses as our approach
which, they claim, explains the observed long--ranged meniscus deformation.
Here we would like to note two important errors which flaw their
analysis:
\begin{enumerate}
\item Eq.~(3.15) in Ref.~\cite{Kra04}, in terms of which they
  interpret their observations, can be obtained by inserting the
  large--distance ($r \gg r_{0, \rm ref}$) asymptotic behavior of
  $\Pi(r)$ in our Eq.~(\ref{eq:sol1}) (equivalent to their Eq.~(3.14)
  up to an additive constant) both into the integral term {\em and into
    the definition of $\varepsilon_\Pi$} (Eq.~(\ref{eq:epsPi})). Since
  the dominant contribution to the integral defining $\varepsilon_\Pi$
  stems from points $r \simeq r_{0,\rm ref}$, this procedure is clearly
  inadmissible. As a consequence, they obtain a wrong, nonvanishing
  prefactor of the logarithm, in spite of their explicit consideration
  of mechanical isolation (see their Eq.~(6.6)).
\item In order to calculate the intercolloidal effective potential within
  the superposition approximation, the formula relating $V_{\rm
    men}(d)$ to the meniscus deformation $u(d)$ {\em in the presence
    of gravity alone} (i.e., Eq.~(\ref{eq:flotV})) is applied even though $\Pi \neq 0$. 
  Thus, an additional term contributing to the leading logarithm is not included
  (compare  their Eq.~(3.16) and our Eq.~(\ref{eq:logV})).
\end{enumerate}
In Ref.~\cite{Kra04} also a numerical analysis is carried out. A
detailed study 
of the relation between the results of this numerical analysis and our theoretical predictions
will be published elsewhere~\cite{ODDDKB04}.

\bigskip

It may be possible that the presence of an external field is consistent with the data
from Refs.~\cite{Nik02, Kra04},
as well as with the presence of the secondary
potential minimum observed in the experiments using planar troughs, in
particular in the cases $d_{\rm min}>10 R$, which fall into the intermediate
asymptotic regime considered here. But this still remains  as an open
problem.

\subsection{Outlook}

Given that already nanometer distortions of the meniscus produce noticeable
attractions, the surface topography of colloids might be relevant.
In Ref.~\cite{Sta00}, colloidal surface roughness is proposed as an
explanation of the attraction. The meniscus contact line is assumed to
be pinned at defects on the colloid surface caused by surface roughness.
This imposes a different boundary condition for the meniscus at
contact, which is then deformed even in the absence of electrostatic
forces.
The corresponding analysis in Ref.~\cite{Sta00} is concerned only with the term $\propto
(\nabla {u}_1) \cdot (\nabla {u}_2)$ in the expression~(\ref{eq:Vmen1}),
leading to the conclusion that $V_{\rm men}(d)$ decays as $d^{-4}$ and
corresponds to an attractive potential with a strength of $10^4 \, k_B
T$ for meniscus deformations of the order of $50 \, {\rm nm}$.
This conclusion, however, cannot be simply carried over to the case
of charged colloids. As we have shown the contributions of
the terms $\propto \Pi_1 u_2$ and $\propto F \bar{u}$ in
Eq.~(\ref{eq:Vmen1}) are relevant and can change the qualitative
behavior of $V_{\rm men}(d)$ even in the limit of point--colloids. 
It would be worthwhile to generalize the analysis of Ref.~\cite{Sta00}
along the line of arguments presented here in order to assess the importance of surface
roughness. This should be complemented by more precise experimental
information about the actual colloidal topography.

Recently, an explanation based on a contaminated interface has been
advocated~\cite{FMMH04}. The air--water interface would be actually a
two--dimensional emulsion consisting of hydrophylic (water) patches and
hydrophobic (silicon oil) patches.  The colloidal particles
(hydrophobic in character according to Ref.~\cite{FMMH04}) would
cluster in the hydrophobic patches. Thus, confinement of the colloids
by finite--size hydrophobic patches would give the impression of an
effective intercolloidal attraction. At present, this explanation is
only of qualitative nature.

Thermal fluctuations of the interface position around its mean value
$u({\bf r})$ also induce an effective interaction between the colloids
which confine these fluctuations (Casimir--like force).
Using a Gaussian model of the fluctuating interface in analogy to the
procedure employed in Ref.~\cite{Gol96} for calculating
fluctuation--induced forces between rods in a membrane, one finds 
for uncharged colloids a
fluctuation--induced potential $V_{\rm fluc} \simeq -k_B T\; (r_{0, \rm
ref}/d)^4$, which is too small and falls off more rapidly than the
intercolloidal dipole repulsion. Here, the generalization to the charged case
might give hints for the effective attractions between {\em very}
small particles trapped at interfaces. 
Concerning particle sizes well below the Debye length, one
should modify our analysis to account for the overlap of the screening
ionic clouds (this would affect, e.g., the superposition approximation
for the stress field $\Pi$, Eq.~(\ref{eq:superposition})).


Finally, in Ref.~\cite{SCLN04} the attraction of particles trapped at
a nematic--air interface is reported and an explanation in terms of a
logarithmic meniscus deformation is proposed which parallels the
explanation given  in Ref.~\cite{Nik02}: in this case, the deformation would
be caused by the elastic distortion induced by the particles on the
nematic phase~\cite{SCLN04}. Our detailed theoretical treatment shows
that no long--range (logarithmic) meniscus distortion can arise on an
interface if the system is mechanically isolated and the excess free energy
of the perturbed interface is correctly described by an expression like
Eq.~(\ref{eq:F}).
Thus it appears that the simple explanation of the observed colloidal 
attractions given in Ref.~\cite{SCLN04} is not correct.
However, it is not clear whether the free energy of a distorted 
nematic--air interface is equivalent to that of a simple fluid interface due
to the long--ranged interactions  in the nematic bulk caused by defects.
 Thus, a
generalization of our theory to interfaces involving nematic phases
 would be desirable in
order to assess the possibility of long--ranged  colloidal attractions
in more detail.

\section{Summary}
\label{sec:summary}

We have analyzed the effective force  induced by capillary deformation
between two smooth spherical colloids floating at a fluid interface.
The relevant degrees of freedom are the meniscus
deformation and the height of the colloids (Fig.~\ref{fig:pert}), 
whose equilibrium values
are given by minimization of a free energy functional. This functional
was derived assuming small deviations from a reference configuration (Fig.~\ref{fig:ref}).
It incorporates the surface tensions of the  three
interfaces involved (``colloid -- fluid phase 1'', ``colloid -- fluid phase 2'',
and ``fluid phase 1 -- fluid phase 2''), the potential energy of the
colloids under the action of a force $F$, the potential energy of the
fluid interface in an arbitrary surface stress field $\Pi({\bf r})$,
and the potential energy due to a restoring force acting on the
interface (Eqs.~(\ref{eq:Fdef})--(\ref{eq:F})). 
The effective intercolloidal potential (Eq.~(\ref{eq:Veff_a})) was calculated in the
limit of large separations by using the superposition
approximation (Eq.~(\ref{eq:superposition})). 
We have shown in this limit that the contribution to the
effective potential by the interfaces ``colloid -- fluid phases'' is subdominant.  If
the total force acting on the system, consisting of the two colloids
and the meniscus,  does not
vanish, the presence of the restoring force acting on the fluid interface 
is essential ---
although its precise form does not matter (Subsec.~\ref{sec:lambda}). In this case, 
the effective
interaction is long--ranged and attractive (Eq.~(\ref{eq:logV})). If the total force
vanishes, the restoring force is irrelevant altogether, the effective
interaction is shorter--ranged (Eq.~(\ref{eq:shortrangeV})), 
and it cannot be computed reliably within
the superposition approximation.

As an application, we have considered the case of like--charged,
micrometer--sized particles when the capillary deformation is due to
the ensuing electrostatic field.
We have discussed how one can tune the
long--ranged attractive interaction by an external electric field, 
but we conclude that the
experimentally observed attraction in an isolated system cannot be
explained within the present model. Possible directions for future research
such as colloidal surface roughness and fluctuations of the interface
have been discussed.

\begin{acknowledgments}
  We kindly acknowledge helpful discussions on the subject of the
  manuscript with M.~Nikolaides, A.~Bausch, K.~Danov, and
  P.~Kralchevsky.
\end{acknowledgments}

\appendix

\section{Free energy of colloid--fluid contact}
\label{sec:Fcont}

\begin{figure}
  \begin{center}
    \epsfig{file=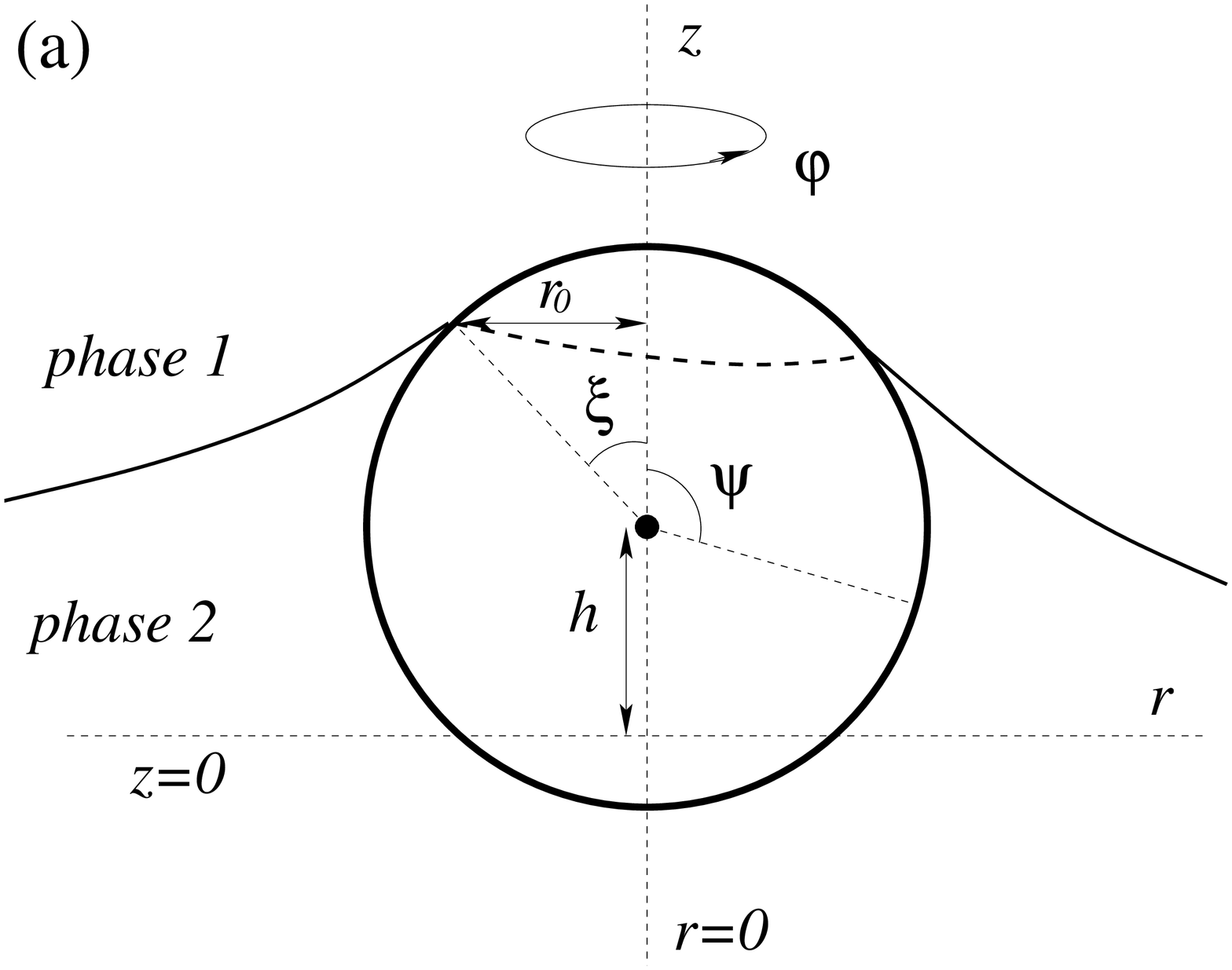,width=.5\textwidth}
    \hspace{0.5cm}
    \epsfig{file=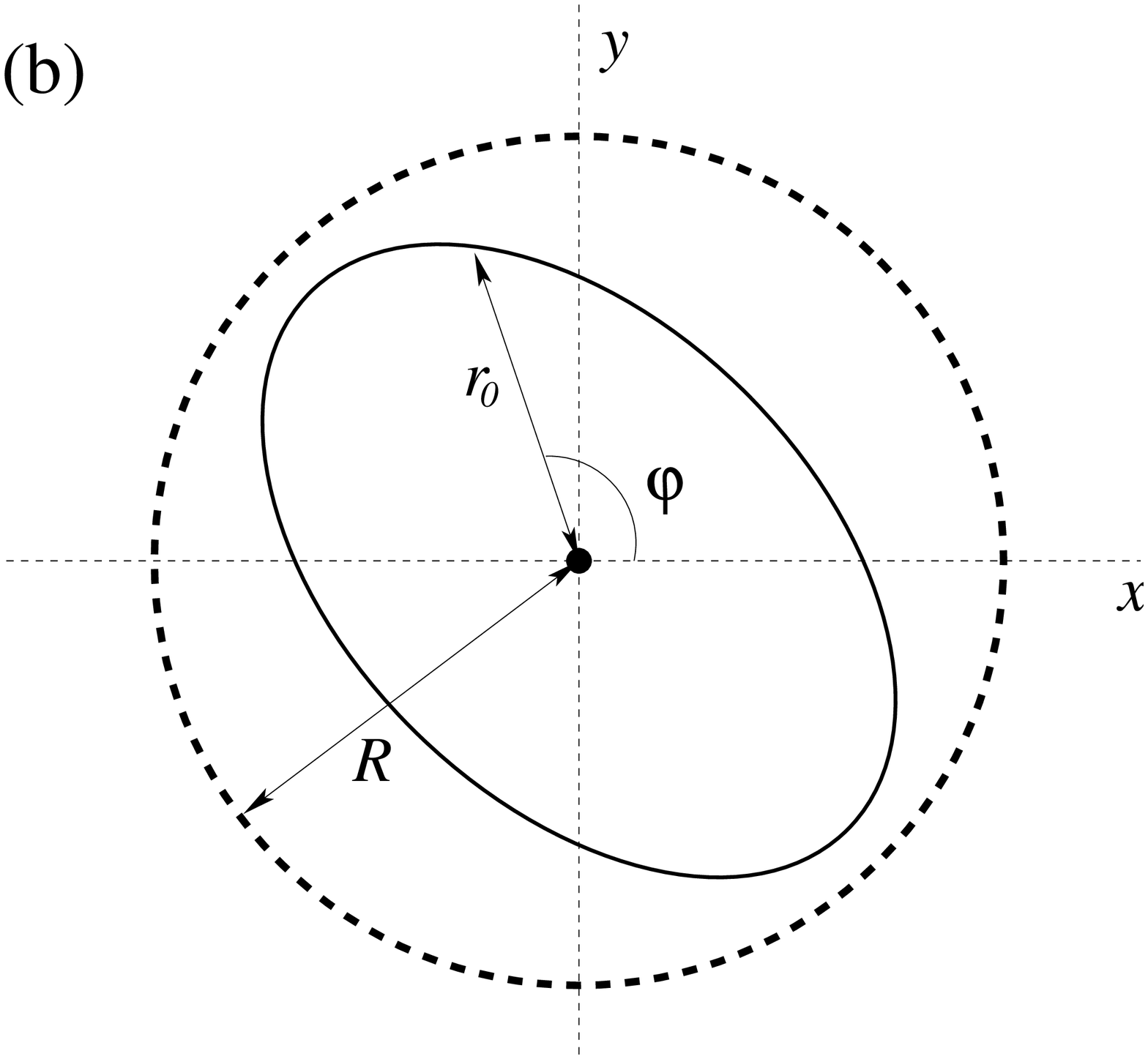,width=.4\textwidth}
    \caption{
      Geometric description of the three--phase contact line between the
    fluid--fluid interface and the surface of the colloid. Side view (a): 
      The auxiliary variables
      $r_0(\varphi)$ and $\xi(\varphi)$ in general depend on the angle of
      revolution $\varphi$. $\psi$ denotes the polar angle of points on the
      colloid spherical surface. Top view (b): The dashed
      line is the circumference of radius $R$ of the spherical colloid. The
      solid line is the projection of a noncircular contact line.}
    \label{fig:contactline}
  \end{center}
\end{figure}

In this appendix we determine the contribution to the free energy due
to the exposure of the colloid to the two fluid phases. Here we consider
the general case of no rotational invariance, so that the
meniscus height $u(r,\varphi)$ depends on the distance $r$ from the axis
through the colloid center and on the angle of revolution $\varphi$
around this axis (Fig.~\ref{fig:contactline}). Accordingly, the auxiliary variables
$r_0(\varphi)$ and $\xi(\varphi)$ also depend on the revolution angle
and one has
\begin{equation}
  \label{eq:r0}
  r_0 (\varphi) = R \sin \xi(\varphi) 
\end{equation}
and
\begin{equation}
  \label{eq:zR}
  h = u(r_0(\varphi), \varphi) - R \cos \xi(\varphi) ,
\end{equation}
where $u(r_0(\varphi), \varphi)$ is the meniscus height at contact.
The surface areas of the colloid in contact with phase 1 and phase 2
are
\begin{equation}
  \label{eq:A1}
  A_1 = R^ 2 \int_0^{2 \pi} \!\!\!\! d \varphi \int_0^{\xi(\varphi)} \!\!\!\! d \psi \, \sin \psi = 
  R^ 2 \int_0^{2 \pi} \!\!\!\! d \varphi \, [1 - \cos \xi(\varphi) ] 
\end{equation}
and
\begin{equation}
  \label{eq:A2}
  A_2 = 4 \pi R^2 - A_1 ,
\end{equation}
respectively.
The expression for the free energy in Eq.~(\ref{eq:Fcont}) is
based on these formulae for the special case $\xi_{\rm ref}(\varphi) = \theta$
(Fig.~\ref{fig:ref}) and on  Young's law (Eq.~(\ref{eq:young})):
\begin{eqnarray}
  {\cal F}_{\rm cont} & = & \gamma R^2 \int_0^{2 \pi} \!\!\!\! d\varphi \, 
        [\cos^2 \theta - \cos \theta \cos \xi(\varphi)] = \nonumber \\
& = & \frac{1}{2} \gamma R^2 \int_0^{2 \pi} \!\!\!\! d \varphi \, 
[\cos \xi(\varphi) - \cos \theta]^2 +  
\frac{1}{2} \gamma R^2 \int_0^{2 \pi} \!\!\!\! d \varphi \, 
[\cos^2 \theta - \cos^2 \xi(\varphi)] \nonumber \\
& = & \frac{1}{2} \gamma \int_0^{2 \pi} \!\!\!\! d \varphi \, 
         [u(r_0(\varphi), \varphi) - \Delta h]^2 +
  \frac{1}{2} \gamma \int_0^{2 \pi} \!\!\!\! d \varphi \, 
[r_0^2 (\varphi) - r_{0,\rm ref}^2 ] .
\end{eqnarray}
The second term, which arises upon completing the square, represents
the change of the meniscus area which is cut out by the colloid.
Since $u$ and $\Delta h$ are already of first order in $\varepsilon_\Pi$ or
$\varepsilon_F$, one can replace $r_0(\varphi)$ by $r_{0, \rm ref}$ in
the first term and obtains 
\begin{equation}
  \label{eq:almostFcont_aniso}
  {\cal F}_{\rm cont} \simeq \frac{1}{2} \gamma 
  \int_0^{2 \pi} \!\!\!\! d \varphi \, [u(r_{0,\rm ref}, \varphi) -
  \Delta h]^2 +
\frac{1}{2} \gamma \int_0^{2 \pi} \!\!\!\! d \varphi \, 
[r_0^2 (\varphi) - r_{0,\rm ref}^2] ,
\end{equation}
plus corrections of at least third order in $\varepsilon_\Pi$ or
$\varepsilon_F$. In the special case of rotational invariance, this
expression reduces to Eq.~(\ref{eq:Fcont_iso}). For the purpose of
Sec.~\ref{sec:2coll} one can rewrite Eq.~(\ref{eq:almostFcont_aniso}) as the
line integral 
\begin{equation}
  \label{eq:Fcont_aniso}
  {\cal F}_{\rm cont} \simeq \frac{\gamma}{2 r_{0,\rm ref}}
  \oint_{\partial S} \!\!\! d \ell \, \left\{ [u - \Delta h]^2 +
[r_0^2 - r_{0,\rm ref}^2] \right\} ,
\end{equation}
where $S$ is a circular disk of radius $r_{0,\rm ref}$, here centered
at the origin, and $\partial S$ denotes its circumference.

\section{Deformation of a spherical reference interface}
\label{sec:droplet}

In this appendix we calculate the deformation of the interface of a
spherical droplet due to a floating colloidal particle.
The reference
configuration corresponds to a colloid floating at the surface of the
spherical droplet such that Eq.~(\ref{eq:young}) is fulfilled. Here we
will consider only axially symmetric configurations (see
Fig.~\ref{fig:droplet}). In terms of the polar angle $\psi \in  [0, \pi]$,
the distance $r$ measured along the
unperturbed spherical droplet surface is $r= R_{\rm drop} \psi$. 
(Note that in this
appendix we consider $u(\cdot)$ and $\Pi(\cdot)$ as functions of
$\psi$ instead of $r$.) We define $\psi_0 := r_{0, \rm ref}/R_{\rm drop}$
and we introduce the angle $\psi_1$ with  $\pi> \psi_1 \gg \psi_0$ at which
another boundary condition (to be specified below) holds reflecting the
fact that the droplet is fixed in space. With this additional
boundary condition we model closely  the actual experimental setup
of Ref.~\cite{Nik02}, where the droplet was attached to a glass plate with a 
contact angle less than $\pi$ \cite{Nik04}. 

For the present case the free energy given by Eq.~(\ref{eq:F}) must
be modified in several ways:
\begin{itemize}
\item In spherical coordinates, ${\cal F}_{\rm men}$ reads
  \begin{equation}
    {\cal F}_{\rm men} = \gamma \int_{S_{\rm men}} \!\!\!\!\!\! dA \; (R_{\rm drop} + u)^2
    \sqrt{1+\left|\frac{\nabla u}{R_{\rm drop} + u}\right|^2} - 
    \gamma \int_{S_{\rm men,ref}} \!\!\!\!\!\! dA \; R_{\rm drop}^2 \, ,
  \end{equation}
  where $dA = 2 \pi \sin \psi \, d\psi$ is the differential of the solid angle,
  and $\nabla={\bf e}_{\psi} (d/d\psi)$ is the gradient operator on the
  unit sphere. As before, this functional is now expanded for small
  deviations from the reference configuration, and we obtain
  \begin{eqnarray}
    \label{eq:fmen1}
    {\cal F}_{\rm men} 
    & \simeq & \gamma \pi (r_{0,\rm ref}^2 - r_0^2) +
    2 \pi \gamma \int_{\psi_{0}}^{\psi_1} \!\!\! d\psi \; \sin\psi \left[ 
      \frac{1}{2} \left(\frac{d u}{d \psi} \right)^2 + u^2 + 2 R_{\rm drop} u \right] ,
  \end{eqnarray}
  where we have neglected terms evaluated at $\psi_0$ which vanish in the limit $R_{\rm drop} \to \infty$.
  
\item The term ${\cal F}_{\rm inter}$ in Eq.~(\ref{eq:Finter}) has to
  be amended by the extra work upon deformation which is caused by the pressure
  difference in the reference configuration between the interior and the exterior of the droplet, 
  $\Delta P = 2 \gamma/R_{\rm drop}$:
  \begin{eqnarray}
    {\cal F}_{\rm inter} &=&- \int_{S_{\rm men}} \!\!\!\!\!\! dA \; \left[
    R_{\rm drop}^2\,\Pi\;u + (\Delta P)  \int_{0}^u \!\!\! d\tilde{u} \; 
    (R_{\rm drop} + \tilde{u})^2 \right] 
     \nonumber \\ 
  \label{eq:finter1}
   & \simeq &- 2 \pi \gamma \int_{\psi_{0}}^{\psi_1} \!\!\! d\psi \; \sin\psi \, 
    \left[ \frac{R_{\rm drop}^2}{\gamma} \Pi \, u + (2 R_{\rm drop} u + 2 u^2)\right] .
  \end{eqnarray}
  Since $\Delta P$ is of order $\varepsilon^0$, we have to keep 
  also terms $\propto u^2$ in this expression for consistency. For
  simplicity we have neglected the contribution due to the change of
  $S_{\rm men}$ during the deformation. Such a  term would affect the
  behavior of the meniscus  only near $\psi_0$ and vanishes in the limit
  $R_{\rm drop} \to \infty$.
  
\item The change in volume of the perturbed droplet reads
  \begin{eqnarray}
    \Delta V = \int_{S_{\rm men}} \!\!\!\!\!\! dA \; \frac{1}{3}(R_{\rm drop} + u)^3 -
    \int_{S_{\rm men, ref}} \!\!\!\!\!\! dA \; \frac{1}{3} R_{\rm drop}^3 +
    \delta
  \end{eqnarray}
  where $\delta$ is a contribution due to the spherical shape of the
  colloid; it depends on $\Delta h$ and $\Delta r_0$ but can be safely neglected.
  (This is the same kind of term encountered in calculating
  ${\cal F}_{\rm vol}$, see Eq.~(\ref{eq:Fvol}).) Expanding
   $\Delta V$ for small deformations
  $u$ up to linear order yields
  \begin{equation}
    \label{eq:deltaV}
    \Delta V \simeq 2 \pi R_{\rm drop}^2 \int_{\psi_{0}}^{\psi_1} \!\!\!\! d\psi \,
    \sin\psi \; u .
  \end{equation}
  Physically, the deformation of the interface occurs under the constraint
  that the droplet volume remains unchanged. This constraint ($\Delta V = 0$)
  will be implemented in the free energy
  functional by means of a Lagrange multiplier $\mu$ which itself turns out to be
  linear in the parameters $\varepsilon_\Pi$ and $\varepsilon_F$. This justifies
  keeping only the linear term in Eq.~(\ref{eq:deltaV}).
\end{itemize}
The contribution ${\cal F}_{\rm cont}$  
(see Eq.~(\ref{eq:Fcont_iso}))  is
concerned only with quantities at $\psi_0$, the colloid surface, and thus
in the limit of large droplet radii this contribution are the same as in the planar case.
The free energy related to the work done upon moving the colloid,
${\cal F}_{\rm coll}$ (see Eq.~(\ref{eq:Fcoll})), remains also unchanged.
For simplicity we also neglect gravity and set
$\lambda^{-1}=0$. In conclusion, the free energy functional to be minimized
is the sum ${\cal F}_{\rm cont} + {\cal F}_{\rm men} 
  + {\cal F}_{\rm inter} + {\cal F}_{\rm coll}$ plus the 
  constraint term $(2\pi\gamma/R_{\rm drop})\;\mu\;\Delta V$ 
(where $\mu$ is a dimensionless Lagrange multiplier):
\begin{equation}
  {\cal F} =
  2 \pi \gamma \int_{\psi_0}^{\psi_1} d\psi \; \sin \psi  
  \left[ \frac{1}{2}\left(\frac{d u}{d \psi} \right)^2 - u^2 + \mu R_{\rm drop} u
    - \frac{R_{\rm drop}^2}{\gamma} \Pi \, u \right] +
  \pi \gamma [ u_0 - \Delta h ]^2 -
  F \Delta h .
\end{equation}
The quantity $-
\gamma \mu/R_{\rm drop}$ can be interpreted as  a homogeneous pressure field enforcing
the constant--volume constraint.
We note that the terms linear in $u$ present in Eqs.~(\ref{eq:fmen1}) and
(\ref{eq:finter1}) have cancelled in the total free energy.
Minimization with respect to $\Delta h$ yields the same result as in
Eq.~(\ref{eq:h_eq}). Subsequent minimization with respect to $u(\psi)$
leads to
\begin{equation}
  \label{eq:spherYL}
  \frac{1}{\sin \psi} \frac{d}{d \psi} \left( \sin\psi \frac{d u}{d \psi} \right) + 2 u =
  - \frac{R_{\rm drop}^2}{\gamma} \Pi + \mu R_{\rm drop}
\end{equation}
and (compare with Eq.~(\ref{eq:u'0}))
\begin{equation}
  \label{eq:u_0}
  \left. \frac{d u}{d \psi} \right|_{\psi_0} = \varepsilon_F R_{\rm drop} \, .
\end{equation}
The solution must satisfy the constant--volume
constraint
\begin{equation}
  \label{eq:constraint}
  \int_{\psi_0}^{\psi_1} \!\!\!\! d\psi \, \sin \psi \; u(\psi) = 0 .
\end{equation}
Finally, the second boundary condition mentioned at the beginning expresses
the physical requirement that the droplet is fixed in space.
(Otherwise, the application of a localized force at the interface would  shift
the droplet as a whole without deforming it.) As an example, we assume
that the droplet interface is pinned at $\psi_1$ (another example, force balance
by suitable localized counterstresses, is
studied in Ref.~\cite{MoWi93}):
\begin{equation}
  \label{eq:u_1}
  u(\psi=\psi_1) = 0. 
\end{equation}

The general solution of the inhomogeneous Legendre equation (\ref{eq:spherYL}) is
\begin{equation}
  \label{eq:sol_drop}
  u(\psi) = A P(\psi) + B Q(\psi) + \frac{1}{2} \mu R_{\rm drop} + 
  \frac{R_{\rm drop}^2}{\gamma} \int_{\psi}^{\psi_1} \!\!\!\! d\sigma \; \sin \sigma \, 
  [P(\sigma) Q(\psi) - P(\psi) Q(\sigma)] \Pi(\sigma) ,
\end{equation}
where $A$ and $B$ are integration constants and 
\begin{eqnarray}
  P(\psi) & := & \cos \psi , \nonumber \\
  Q(\psi) & := & 1 + \cos \psi \, \ln \tan \frac{\psi}{2} \;,
\end{eqnarray}
are solutions of the homogeneous Legendre equation.
For notational simplicity, evaluation of a function at
$\psi_0$($\psi_1$) will be denoted by the subscript $_{0 (1)}$.

From Eq.~(\ref{eq:u_1}) one obtains
\begin{equation}
  \label{eq:mu}
  \frac{1}{2} \mu R_{\rm drop} = - A P_1 - B Q_1 .
\end{equation}
From the other boundary condition~(\ref{eq:u_0}) it follows that
\begin{equation}
  \label{eq:ab1}
  \varepsilon_F R_{\rm drop} = A P'_0 + B Q'_0 + 
  \frac{R_{\rm drop}^2}{\gamma} \int_{\psi_0}^{\psi_1} \!\!\!\! d\sigma \; \sin \sigma \, 
  W(\sigma) \Pi(\sigma) ,
\end{equation}
where we have defined the auxiliary function
\begin{equation}
  W(\sigma) := [P(\sigma) Q'_0 - P'_0 Q(\sigma)] . 
\end{equation}
Finally, in order to impose the integral constraint~(\ref{eq:constraint}) we
employ the following identities:
\begin{eqnarray}
  \int_{\psi_0}^{\psi_1} \!\!\!\! d\psi \, \sin \psi \; \int_{\psi}^{\psi_1} \!\!\!\! d\sigma \; \sin \sigma \, 
  [P(\sigma) Q(\psi) - P(\psi) Q(\sigma)] \Pi(\sigma) =  \\
  \int_{\psi_0}^{\psi_1} \!\!\!\! d\sigma \, \sin \sigma \; \Pi(\sigma) \; \int_{\psi_0}^{\sigma} \!\!\!\! d\psi \; \sin \psi \, 
  [P(\sigma) Q(\psi) - P(\psi) Q(\sigma)] = \nonumber \\
  \mbox{} - \int_{\psi_0}^{\psi_1} \!\!\!\! d\sigma \, \sin \sigma \; \Pi(\sigma) \; \frac{1}{2} \left[1-\frac{W(\sigma)}{W_0}\right] \;. \nonumber 
\end{eqnarray}
The first equality follows from interchanging the order of integration.
The last equality follows most easily by noticing (i) that 
the $\psi$--integral in the
second line is, as a function of $\sigma$, a solution of the differential
equation~(\ref{eq:spherYL}) with $u$ replaced by this integral, $\mu R_{\rm drop}$ 
replaced by $-1$, $\Pi=0$, and
the analogues of the boundary conditions $u_0=0=u'_0$, and (ii) that the function
$(W(\sigma)/W_0 -1)/2$ is the solution to the same differential equation
with the same boundary conditions.
With these identities Eq.~(\ref{eq:constraint}) turns into
\begin{equation}
  \label{eq:ab2}
  0 = A {\cal P} + B {\cal Q} -  \frac{1}{2} \int_{\psi_0}^{\psi_1} \!\!\!\! d\sigma \, \sin \sigma \; \Pi(\sigma) \; \left[1-\frac{W(\sigma)}{W_0}\right] \;, 
\end{equation}
where we have introduced the numerical constants
\begin{eqnarray}
  {\cal P} & := & \int_{\psi_0}^{\psi_1} \!\!\!\! d\psi \, \sin \psi \; [P(\psi) - P_1] \nonumber \\
  {\cal Q} & := & \int_{\psi_0}^{\psi_1} \!\!\!\! d\psi \, \sin \psi \; [Q(\psi) - Q_1] .
\end{eqnarray}
We note that the constants $A$ and $B$ are determined by the two linear 
Eqs.~(\ref{eq:ab1}) and (\ref{eq:ab2}).

These results permit us to study the intermediate asymptotic regime
$r_{0, \rm ref}, r \ll R_{\rm drop}$. 
In the expressions derived above, we have to take the limit
$R_{\rm drop} \to \infty$ while keeping fixed $\psi_1$, $r=R_{\rm drop} \psi$ 
and $r_{0, \rm ref}=R_{\rm drop} \psi_0$ (so that $\psi_0, \psi \ll 1$).
In this limit, the
constants ${\cal P}$ and ${\cal Q}$ are finite, while $P(\psi) \simeq
1$ and $Q(\psi) \simeq \ln(r/R_{\rm drop}) - \ln 2 + 1$. The
integral over $\Pi$ in Eq.~(\ref{eq:ab1}) simplifies as follows
(provided that the stress field $\Pi(\psi)$ decays sufficiently
fast):
\begin{eqnarray}
  \int_{\psi_0}^{\psi_1} \!\!\!\! d\sigma \; \sin \sigma \, W(\sigma) \Pi(\sigma) & = & 
  \int_{r_0}^{\psi_1 R_{\rm drop}} \!\!\!\! ds \; \sin \left(\frac{s}{R_{\rm drop}}\right) \, 
  W\left(\frac{s}{R_{\rm drop}}\right) \Pi\left(\frac{s}{R_{\rm drop}}\right) \nonumber \\
  & \simeq & \frac{1}{R_{\rm drop}^2} \left(\frac{R_{\rm drop}}{r_0}\right) 
  \int_{r_0}^{+\infty} \!\!\!\! ds \; s \, \Pi\left(\frac{s}{R_{\rm drop}}\right) 
\end{eqnarray}
where we have used $W(s/R_{\rm drop}) \simeq R_{\rm drop}/r_0$.
The integrals over $\Pi$ in Eqs.~(\ref{eq:ab2}) and
(\ref{eq:sol_drop}) can be simplified in the same manner. From
Eq.~(\ref{eq:ab1}) we obtain
\begin{equation}
  B \simeq r_0 (\varepsilon_F - \varepsilon_\Pi) , 
\end{equation}
and Eq.~(\ref{eq:ab2}) yields
\begin{equation}
  A \simeq - \frac{\cal Q}{\cal P} r_0 (\varepsilon_F - \varepsilon_\Pi) .
\end{equation}
One finds indeed that the Lagrange multiplier $\mu$ determined by
Eq.~(\ref{eq:mu}) is linear in $\varepsilon_F - \varepsilon_\Pi$.
Finally, the general solution~(\ref{eq:sol_drop}) simplifies to
\begin{equation}
  u \simeq A (1-P_1) + B \left( \ln\frac{r}{R_{\rm drop}} - \ln 2 + 1 - Q_1 \right) 
  - \frac{1}{\gamma} \int_{r}^{+\infty} \!\!\!\! ds \; s \, \Pi\left(\frac{s}{R_{\rm drop}}\right)
  \ln \frac{s}{r} ,
\end{equation}
which renders Eq.~(\ref{eq:sol3}) with the constant
\begin{equation}
  \label{eq:tildeC}
  \tilde{C} = 2 \exp{\left[ (1-P_1) \frac{\cal Q}{\cal P} + Q_1 -1 \right]} .
\end{equation}
This constant is finite and non--vanishing provided $\psi_1$ is not close to $\pi$. 
The case $\psi_1=\pi$ is pathological because $u' (\pi) \equiv 0$ for a
smooth profile, so that the boundary condition~(\ref{eq:u_1}) would
overdetermine the problem.

\end{document}